\documentclass[intlimits,twoside,a4paper]{article}

\usepackage{bm}
\usepackage[T2A]{fontenc}
\usepackage[cp1251]{inputenc}

\usepackage[eqsecnum]{cmpj3}

\issue{2018}{21}{4}{43502}
\doinumber{10.5488/CMP.21.43502}

\title[The EoS of a supercritical cell fluid model]%
{The equation of state of a cell fluid model in the supercritical region}

\author{M.P. Kozlovskii, I.V. Pylyuk, O.A. Dobush}
\address{Institute for Condensed Matter Physics of the National Academy of Sciences of Ukraine, \\ 1 Svientsitskii St., 79011 Lviv, Ukraine}

\newcommand{\non}{\nonumber \\}

\newcommand{\be}{\begin{equation}}
\newcommand{\ee}{\end{equation}}
\newcommand{\bea}{\begin{eqnarray}}
\newcommand{\eea}{\end{eqnarray}}
\newcommand{\sli}{\sum\limits}

\newcommand{\lp}{\left (}
\newcommand{\rp}{\right )}

\newcommand{\ld}{\left .}
\newcommand{\rdot}{\right .}

\newcommand{\vk}{\vec{k}}
\newcommand{\vl}{\vec{l}}

\newcommand{\cB}{{\mathcal{B}}}

\newcommand{\rhok}{{\rho_{\vec{k}}}}
\newcommand{\rhomk}{{\rho_{-\vec{k}}}}
\newcommand{\etak}{{\eta_{\vec{k}}}}

\newcommand{\nuk}{{\nu_{\vec{k}}}}

\date{Received May 8, 2018,  in final form October 26, 2018}

\begin{document}

\maketitle

\begin{abstract}
The analytic method for deriving the equation of state of a cell
fluid model in the region above the critical temperature ($T \geqslant T_\text{c}$) is
elaborated using the renormalization group transformation in
the collective variables set.
Mathematical description with allowance for non-Gaussian fluctuations of the order parameter is
performed in the vicinity of the critical point on the basis of
the $\rho^4$ model. The proposed method of calculation of the grand partition function allows one to obtain the equation for the critical temperature of the fluid model in addition to universal quantities such as critical exponents of the correlation length. The isothermal compressibility is plotted as a function of density. The line of extrema of the compressibility in the supercritical region is also represented.


\pacs 51.30.+i, 64.60.fd
\end{abstract}

\section{Introduction}
\label{intro}

In recent decades the interest in supercritical fluids is steadily increasing. They possess unique properties which appear to be efficient for chemical, technological and industrial appliances \cite{Marr_2000}. The behavior of supercritical fluids is intensively investigated in experiments by methods of computer simulations and by development of theoretical approaches. These directions of research are well covered in recent reviews by Yoon, Lee~\cite{Yoon_2017} and Vega~\cite{Vega}. The theoretical approaches for description of the critical phenomena in fluids are represented by the statistical mechanical theories such as the theory of integral equations, which consists of homogeneous Ornstein-Zernike Equation~\cite{Hansen,Kovalenko} and the closure relations such as the Percus-Yevick~\cite{PY} and hypernetted chain closures~\cite{Hirata_2003}, and the fluctuation theory of solutions~\cite{Smith_2013}. The renormalization group (RG) theory~\cite{wilson} has also been very successful in describing the properties of systems near their critical point.
 The conception of RG theory was used by Yukhnovskii~\cite{Yu_2018} to make a total integration in the partition function of the Ising model and the grand partition function of a gas-liquid system in the phase-space of collective variables and to investigate the behavior of these systems in the vicinity of a critical point.

For over a century, scientists are trying to theoretically describe the nature of phase transitions and critical phenomena in liquid systems. In the course of long years of strenuous work, various worthy approaches have been elaborated to solve this problem in the frames of canonical ensemble but the development of resemblant theories on the basis of the grand canonical distribution still remains topical.The latter direction is of particular importance since due to the presence of chemical potential within the framework of the grand canonical ensemble, the actual systems of atoms and molecules can be adequately represented. Only this thermodynamic parameter is responsible for the exchange of constituents between different parts of the system and with the environment. Moreover, it quantitatively describes the tendency of the thermodynamic system to establish a composition equilibrium.

In articles~\cite{kd117,kdp117}, we calculated the grand partition function of a cell fluid model in the mean-field type approximation. In this way, it was possible to describe the first-order phase transition and, in general, the behavior of the system in a wide range of temperatures above and below the critical point, except its vicinity.  In particular, using the Morse interaction potential~\cite{Morse}, which well describes the interaction in liquid alkali metals, we obtained an equation of state, coexistence curves, and values of critical density and critical temperature for sodium and potassium. However, in the mean-field approximation, it is impossible to describe the behavior of a three-dimensional system in a close vicinity of a critical point, where fluctuations play a significant role and collective effects turn out to be primary. The solution of the problem will be the calculation of the grand partition function using the renormalization group transformation.
  We intend to implement the idea of the method elaborated in order to calculate the partition function of the 3D Ising-like model in the external field~\cite{K_2012,KR_2012,K_2009} as well as to analyze the critical behavior of this system. This way makes it possible to provide a thorough study of the critical properties of the correlation function and thermodynamic functions such as specific heat, isothermal compressibility and isobaric expansion, to estimate the values of critical amplitudes and critical parameters, to plot the Widom line~\cite{widom, bryk}, which is necessary for investigating a fluid in the supercritical region. Note that unique properties of the matter appear during phase transitions in other systems. In particular, the articles~\cite{slusar1,slusar2} represent the investigation of the phase transition in a Fermi fluid from a normal state to a state in which the rotational symmetry breaking takes place in the momentum space. The authors obtained analytic expressions for the phase transition temperature as well as the order parameter near the critical point using the equation of self-consistency.

In this article, we propose a theoretical description of the behavior of a cell fluid model near the critical point at temperatures above the critical. Section~\ref{sec2} is devoted to a staged calculation of the grand partition function of the cell fluid model within an approach of collective variables. Furthermore, the recurrence relations between the coefficients of effective non-Gaussian measures of density, the solutions of these relations and the equation for a phase transition temperature are presented. The thermodynamic potential is considered in section~\ref{sec3}. The total expression of the thermodynamic potential in case of temperatures $T>T_\text{c}$ is obtained as a compilation of terms derived for each fluctuation regime. A technique of calculating the equation of state with fluctuational effects taken into account is elaborated in section~\ref{sec4}. The corresponding expressions are derived for the cases of $T \geqslant T_\text{c}$. Plots of the isothermal compressibility and the line based on the points of maxima of isothermal compressibility are also presented in this section. Discussions and conclusions are presented in section~\ref{concl}.

\section{Basic expressions}
\label{sec2}

As in our previous articles~\cite{kd117,kdp117}, the object of the present investigation is a cell fluid model. The cell fluid is an approximation of a continuous system. Under continuous system one should mean the system of volume $V$ composed of $N$ interacting particles. Similarly to the case of a cell gas model~\cite{rebenko}, the idea of the cell fluid~\cite{kkd2018} consists in a fixed partition of the volume $V$ of a system, where $N$ particles reside, on $N_v$ mutually disjoint cubes $\Delta_{\vl} = (-c/2,c/2]^3 \subset \mathbb{R}^3$, each of the volume $v = c^3 = V/N_v$ ($c$
is the side of a cell). Instead of the distance between particles, the distance between the centers of cells is introduced.

It is possible to theoretically describe the behavior of any system having its equation of state. One of the ways to obtain the latter is an analytical calculation of the grand partition function (GPF) of the system. As it had been already shown in~\cite{kd117,kdp117}, the GPF of the cell fluid model within the framework of the grand canonical ensemble is of the form
\be \label{GPF_1}
\Xi =\!\! \sli_{N=0}^{\infty} \! \frac{(z)^N}{N!} \!\! \int \limits_{V} \! (\rd x)^N \!
\exp \! \Bigg[-\frac{\beta}{2} \!\! \sli_{\vec{l}_1,\vec{l}_2\in\Lambda} \!\!\! \tilde U_{l_{12}} \rho_{\vl_1} \! (\eta) \rho_{\vl_2} \! (\eta) \! \Bigg]. \!
\ee
Here and forth, $z = \text{e}^{\beta \mu}$ is the activity, $\beta$ is the inverse temperature, $\mu$ is the chemical potential. Integration over the coordinates of all the particles in the system is noted as $\int \limits_{V} \! (\rd x)^N = \int \limits_{V} \! \rd x_1 \ldots \! \int \limits_{V} \! \rd x_N$, $x_i = \big(x_{i}^{(1)},x_{i}^{(2)}, x_{i}^{(3)}\big) $, $\eta = \{ x_1 , \ldots , x_N \}$ is the set of coordinates. \\
$\tilde U_{l_{12}}\,\, \text{is the potential of interaction}
$, $l_{12}= \big|\vec{l}_{1}- \vec{l}_{2}\big|$ is the difference between two cell vectors.
Each $\vec{l}_i$ belongs to a set $\Lambda$, defined as
\begin{equation*} \label{Lambda}
% \nonumber to remove numbering (before each equation)
\Lambda =\Big\{ \vl = (l_1, l_2, l_3)|l_i = c m_i;\, m_i=1,2,\ldots , N_{a};\, i=1,2,3;~N_v=N_a^3 \Big\}.
\end{equation*}
Here, $N_a$ is the number of cells along each axis, $\rho_{\vl}(\eta)$ is the occupation number of a cell
%equation(1.4)
\begin{equation} \label{loc_dens}
\rho_{\vl}(\eta) = \sli_{x \in \eta} I_{\Delta_{\vl}(x)}\,.
\end{equation}
Here, $I_{\Delta_{\vl}(x)}$ is the indicator of $\Delta_{\vl}$,
that is, $I_{\Delta_{\vl}(x)}=1$ if $x\in \Delta_\ell$ and
$I_{\Delta_{\vl}(x)} =0$ otherwise.
Interaction in the system is expressed as $\tilde U_{l_{12}}$ which is a function of distance between cells. We choose $\tilde U_{l_{12}} = \Psi_{l_{12}} - U_{l_{12}}$ in the form of a lattice analogue of the Morse potential
%equation(2.3)
\begin{equation} \label{Morse_pot}
\Psi_{l_{12}} = D \text{e}^{-2(l_{12}- 1)/\alpha_R}; \quad U_{l_{12}} = 2 D \text{e}^{-(l_{12}- 1)/\alpha_R},
\end{equation}
   here, $\alpha_R = \alpha / R_0$ where $\alpha$ is the effective interaction radius. The parameter $R_0$ corresponds to the minimum of the function $\tilde U_{l_{12}}$ \big[$\tilde U(l_{12}=1)=-D$ determines the depth of the potential well\big].
   Note that in terms of convenience, the $R_0$-units are used for length measuring. In the next sections we present some quantitative results using the following parameters of the interaction potential: $R_0 = 5.3678$ {\AA}, $\alpha_R = 0.3385$, $D = 0.9241 \times 10^{-20}$ J \cite{singh}. The latter parameters are used to describe the interaction in sodium by the Morse potential~\cite{Morse}. The value of the model's parameter $c$ in this case is $c = 1.3424$, which is the same as we used in the mean-field description of the vapor-liquid transition in alkali metals~\cite{kdp117}. The procedure of self-consistent determination of $v = c^3$, as well as of other parameters of the model, is presented in~\cite{kd117,kdp117}.

%sodium (Na).

In our recent publication~\cite{kkd2018} we made an accurate calculation of the grand partition function of a single-sort cell model with Curie-Weiss potential and found that this model has a sequence of first order phase transitions at temperatures below the critical one $T_\text{c}$. The Curie-Weiss potential was chosen in the form of a function
\begin{equation*}
\Phi_{l_{12}} = \left\{ \begin{array}{ll} - J_1/{N_v}\,, & x\in\Delta_{\vl_1}\,,\quad y \in\Delta_{\vl_2}\,, \quad \vl_1\neq \vl_2\,, \\
J_2 \delta_{\vl_1 \vl_2}\,, & x,y\in\Delta_{\vl_1}\,, \quad \vl_1 = \vl_2\,. \end{array} \right.
\end{equation*}
By virtue of the Curie-Weiss approach, it fails to be a function of distance between constituents of the system.
 The first term in $\Phi_{l_{12}}$ with $J_1>0$ describes attraction. It is taken to be equal for all particles. The second term with $J_2>0$ describes the repulsion between two particles contained in one and the same cell.
 In the present research, the potential of interaction is a function of distance. This manoeuvre allows us to take account of the influence of long-range fluctuations. The analogue of interaction $ - J_1  = \text{const}$ is an exponentially decreasing function of distance $\tilde U_{l_{12}} (l_{12} \neq 0) < 0$, the analogue of $J_2$ is $\tilde U_{l_{12}} (l_{12} = 0) > 0$. From this point of view, the constant interaction $\tilde U_{l_{12}} (l_{12} = 0) = D \text{e}^{1/\alpha_R} (\text{e}^{1/\alpha_R} - 2) > 0$ reflects the interaction energy within a cell. Clearly, this energy is of a repulsive nature and corresponds to the interaction between particles located in the same cell.

In~\cite{kd117,kdp117} we obtained a general functional representation of the grand partition function of the cell fluid model in the set of  collective variables in the following form
\begin{equation}\label{GPF_3}
\Xi = \int (\rd \rho)^{N_v} \exp \! \Bigg[ \beta \mu \rho_{0} + \frac{\beta}{2} \! \sum \limits_{\vec{k} \in\mathcal{B}_c} \!\!\! W(k) \rho_{\vk} \rho_{-\vk} \! \Bigg] \!
\prod \limits_{l=1}^{N_v} \left[ \sli_{m=0}^\infty \frac{v^m}{m!} \text{e}^{-pm^2}\delta(\rho_{\vl}-m)\right],
\end{equation}
%\begin{equation}\label{GPF_3}
% \Xi = \! \int \!\! (\rd \rho)^{N_v} \! \exp \! \Bigg[ \beta \mu \rho_{0} + \frac{\beta}{2} \! \sum \limits_{\vec{k} \in\mathcal{B}_c} \!\!\! W(k) \rho_{\vk} \rho_{-\vk} \! \Bigg] \!
% \prod \limits_{l=1}^{N_v} \left[ \sli_{m=0}^\infty \frac{v^m}{m!} \text{e}^{-pm^2}\delta(\rho_{\vl}-m)\right],
%\end{equation}
from which one can see that the occupation numbers of cells $\rho_{\vl}(\eta)$, which are connected with the variables $\rho_{\vl}$, can take values $m = 0, 1, 2, \ldots .$ Due to the term $\text{e}^{-pm^2}$, the more $m$ increases the less $m$-th term contributes to the sum in \eqref{GPF_3}. Therefore, the probability of hosting many particles in a single cell is very small.
Note that
\[ \label{div_ro}
(\rd \rho)^{N_v} = \prod \limits_{\vk\in\cB_c} \rd \rho_{\vk}.
\]
The variable $\rho_{\vk}$ is the representation of $\rho_{\vl}$ in reciprocal space.

Here and forth, the vector $\vec{k}$ belongs to the set
\begin{equation*}\label{Bk}
    \cB_c \! = \! \Big\{  \vk \! = \! (k_{1},k_{2},k_{3}) \, \Big| \, k_{i} \! = \! -\frac{\piup}{c}+\frac{2\piup}{c}\frac{n_{i}}{N_{a}}, \, n_{i} \! = \! 1,2,\ldots,N_{a};\,\, i=1,2,3; \, N_{v} = N_{a}^{3}   \Big\}.
\end{equation*}
The Brillouin zone $ \cB_c $ corresponds to the volume of periodicity $\Lambda$ with cyclic boundary conditions.
The parameter $p$ is expressed as follows:
\begin{equation*}
    p = \frac{1}{2}\beta \chi \Psi (0)\,,
\end{equation*}
and $W(k)$ is the Fourier transform of the effective potential of interaction
\be\label{1d4fa}
W(k) = U(k) - \Psi(k) + \chi \Psi(0)\,.
\ee
The Fourier transforms of the attractive and of the repulsive parts of the Morse potential \eqref{Morse_pot} are, respectively, as follows:
%equation(2.4)
\be \label{fourier_Morse_pot}
U(k) = U(0)  \big( 1 +  \alpha_R^2 k^2 \big)^{-2} \!\!\!\!\!\!\,\,\,\,\,, \qquad \Psi(k) = \Psi(0)  \big( 1 + \alpha_R^2 k^2 /4 \big)^{-2} \!\!\!\!\!\!\,\,\,\,\,,
\ee
where
%equation(2.7)
\be \label{fourier_Morse_pot_0}
\quad U(0) = 16 D \piup\frac{\alpha_R^3}{v} \text{e}^{R_0/\alpha} \quad \text{and} \quad \Psi(0) = D \piup \frac{\alpha_R^3}{v} \text{e}^{2R_0/\alpha}.
\ee
The 
positive parameter $\chi$ forms the Jacobian of transition from individual coordinates to collective variables.

There are at least three possible ways of calculating \eqref{GPF_3}, in order to derive the equation of state.
The first option is to average the potential of interaction in the direct space or to replace it by the one that fails to be a function of the distance between constituents of the system. In this case, one gets a diagonal form in the integrand exponent. We did this in the article~\cite{kkd2018}, using a Curie-Weiss potential to mimic the interaction in a cell model, and rigorously calculated the grand partition function of this system. However, such interaction potential has a non-physical property since the force of interaction is a function of the number of cells. Nevertheless, it allows one to determine sufficient thermodynamic properties.
The second option is to use an approximation of the $\rho^4$-model which consists of the cutting off terms proportional to the fifth power of the variable $\rho_{\vec{k}}$ and higher,
and to use a type of the mean-field approximation considering only variables $\rho_{\vec k}$ with $\vec k = 0$ (see~\cite{kd117,kdp117}).
Both of these options make it possible to explicitly calculate the grand partition function and derive an equation of state. However, using them, one would describe the behavior of the model in a wide range of temperatures excluding a close vicinity of the critical point.

The third option is presented in the present manuscript. It consists in using the idea of the method of calculating the partition function elaborated for a magnet~\cite{ypk102,ypk202}. In this way, it is possible to take into account the influence of long-range fluctuations, which is necessary for the description of critical behavior.
 To follow this purpose, we start from the expression of the grand partition function of the cell fluid in $\rho^4$-model approximation, which is derived from \eqref{GPF_3} (see~\cite{kd117,kdp117})
\be\label{1d1fa}
\Xi = g_W \text{e}^{N_v( E_\mu-a_0)} \int (\rd \rho)^{N_v} \exp \Bigg[ M N_v^{1/2} \rho_0 - \frac{1}{2} \sli_{k\in\cB} d(k) \rhok\rhomk - \frac{a_4}{24} \frac{1}{N_v} \sli_{{k_1,...,k_4}\atop{\vk_i\in\cB}} \rho_{\vk_{1}} ... \rho_{\vk_{4}} \delta_{\vk_{1}+...+\vk_{4}}\Bigg].
\ee
Here,
\bea\label{1d2fa}
&&
g_W = \prod_{\vk\in\cB} \big[2\piup \beta W(k)\big]^{-1/2},\quad a_{34} = - a_3/a_4\,,\non
&&
E_\mu = -\frac{\beta W(0)}{2} (M + \tilde a_1)^2 + M a_{34} + \frac{1}{2}d(0) a_{34}^2 - \frac{a_4}{24} a_{34}^4\,,\non
&&
M = \frac{\mu}{W(0)} - \tilde a_1\,, \quad \tilde a_1 = a_1 + d(0) a_{34} + \frac{a_4}{6} a_{34}^3\,.
\eea
The following expression corresponds to the coefficient $d(k)$
\be\label{1d3fa}
d(k) = \frac{1}{\beta W(k)} - \tilde a_2\,, \quad \tilde a_2 = \frac{a_4}{2}a_{34}^{2} - a_2\,.
\ee
The effective potential of interaction $W(k)$ from \eqref{1d4fa}
is definitely positive when $k=0$ for all $\chi>1$, but for $0<\chi<1$ it is either positive or negative depending on the ratio $R_0/\alpha$.
If $R_0/\alpha = 2.9544$ [which is typical of Na (sodium)], $\chi = 1.1243$ ($p = 1.8100$) and  $v = 2.4191$~\cite{kdp117} we have the following:
\be\label{1d5fa}
a_0= - 0.3350, \quad \tilde a_1 = - 0.1891, \quad \tilde a_2 = 0.3242, \quad a_4 = 0.0376, \quad a_{34} = 2.4925.
\ee
Let us represent $d(k)$ in the form of series in power $k^2$
\be\label{1d7fa}
d(k) = d(0) + 2 b k^2 / \beta W(0) + 0(k^4).
\ee
Here,
\be\label{1d8fa}
b = \alpha_R^2 \big[ U(0) - \Psi(0)/4\big] / W(0) .
\ee
As can be seen from the look of expressions of $d(0)$ \eqref{1d3fa} and $b$ \eqref{1d8fa}, it is expedient to make a change of variables in the expression \eqref{1d1fa}
\be\label{1d9fa}
\rho_{k} = \rho'_k \big[\beta W(0)\big]^{1/2}.
\ee
As a result, we have
\begin{align}
%\bea
\label{1d10fa}
\Xi &= g_W \big[\beta W(0)\big]^{N_v/2}
\text{e}^{N_v( E_\mu-a_0)} \int (\rd \rho)^{N_v} \non
&
\times
\exp \Bigg[ w_0 N_v^{1/2} \rho_0 - \frac{1}{2} \sli_{\vk\in\cB} \big(r_0+2bk^2\big) \rhok\rhomk - \frac{u_0}{24} \frac{1}{N_v} \sli_{\vk_i \in\cB} \rho_{\vk_{1}} ... \rho_{\vk_{4}} \delta_{\vk_{1}+...+\vk_{4}}\Bigg],
%\eea
\end{align}
moreover, (the sign of stress near $\rhok$ is omitted)
\be\label{1d11fa}
w_0=M\big[\beta W(0)\big]^{1/2}, \quad r_0 = 1 - \beta W(0) \tilde a_2\,, \quad u_0 = a_4 \big[\beta W(0)\big]^2.
\ee
Let us make in \eqref{1d10fa} a staged integration keeping with the technique elaborated in~\cite{KPP_2006PA}. One should start from the variables $\rhok$ with large value of the wave vector and end with integration over the variable~$\rho_0$. The latter is the basis of a type of mean-field approximation. The contribution of $\rho_0$ is the main one, in case of temperatures far from $T_\text{c}$ ($T_\text{c}$ is the critical temperature). It is essential to take account of the fluctuations in the vicinity of $T_\text{c}$; thus, one should consider terms $\rhok$ with $k\neq 0$ in calculations.

According to ~\cite{K_2012}, let us write the following:
\be\label{1d12fa}
\sli_{\vk\in\cB} \big(r_0 + 2bk^2\big) \rhok\rhomk = \sli_{\vk\in\cB_1} \big(r_0 + 2 b k^2\big) \rhok\rhomk + \sli_{\vk\in\cB/\cB_1} (r_0+q) \rhok\rhomk\,,
\ee
here, $q = 2bq'$ and $q' = \langle k^2\rangle_{\cB_1 , \cB}$ is the average of $k^2$ on the interval $\vk \in (\cB_1 , \cB )$,
\be\label{1d14fa}
q' = \frac{3}{2} \frac{\piup^2}{c^2} \big(1+s^{-2}\big).
\ee
The range of values $\vk\in\cB_1$ are of the form:
\be\label{1d13fa}
\cB_1 = \left\{ \vk = (k_1, k_2, k_3) | k_i = - \frac{\piup}{c_1} + \frac{2\piup}{c_1} \frac{n_i}{N_{1i}};\,\, n_i = 1,2,...\,, N_{1i};\,\, i = x,y,z;\,\, N_1 = N^3_{1x}\right\}.
\ee
Here, $c_1 = s c$, moreover, the way of dividing the space of collective variables into intervals $(s>1)$ is determined by the parameter $s=B/B_1$.
Rewrite \eqref{1d10fa} as
\begin{align}
%\bea
\label{1d16fa}
\Xi &= G_\mu \int (\rd \rho)^{N_1}  \text{e}^{-\frac{1}{2} \sli_{\vk\in\cB_1}\left[ r_0+2 b k^2 - (r_0+q)\right]  \rhok\rhomk}
\text{e}^{w_0\sqrt{N_v}\rho_0} \prod_{\vk\in\cB_1} \delta(\etak-\rhok)  \non
&
\times
\int (\rd \eta)^{N_v}  \exp \Bigg[ -\frac{1}{2} \sli_{\vk\in\cB} (r_0+q) \eta_{\vk}\eta_{-\vk} -
\frac{u_0}{24} \frac{1}{N_v} \sli_{\vk_{i} \in\cB} \eta_{\vk_{1}}...\eta_{\vk_{4}} \delta_{\vk_{1}+...+\vk_{4}}\Bigg],
%\eea
\end{align}
where
$G_\mu = g_W \big[\beta W(0)\big]^{N_v/2} \text{e}^{N_v( E_\mu-a_0)}$, $N_1=N_v s^{-3}$. Use the integral representation
\[
\delta(\etak - \rhok) = \int (\rd \nu)^{N_1} \exp \Bigg[ 2 \piup \text{i} \sli_{\vk\in\cB_1}\nu_{\vk}\big(\etak - \rhok\big)\Bigg]
\]
and integrate over $N_v$ variables $\etak$. Thus, we have
\begin{align}
%\bea
\label{1d17fa}
\Xi &= G_\mu \big[Q(r_0)\big]^{N_v} j_1 \int (\rd \rho)^{N_1} \prod_{l=1}^{N_1} \int (\rd \nu)^{N_1} \text{e}^{w_0\sqrt{N_v}\rho_0}
\exp \left[ - \frac{1}{2} \sli_{\vk\in\cB_1} 2b\big(k^2-q'\big) \rhok\rhomk \rdot\non
&
- \ld 2 \piup \text{i} \nu_{\vl} \rho_{\vl} - \sum_{m=1}^{m_0} (2\piup)^{2m} \frac{P_{2m}}{(2m)!} \nu_{\vl}^{2m}\right].
%\eea
\end{align}
Here, $\nu_{\vl} = \frac{1}{\sqrt{N_1}} \sli_{\vk\in\cB_1}\nu_{\vk} \text{e}^{\text{i}\vk\vl}$ is a site representation of the variable $\nuk$,
$j_1 = 2^{(N_1-1)/2}$ is the Jacobian of transition from variables $\nu_{\vk}$ to $\nu_{\vl}$\,,
\be\label{1d18fa}
Q(r_0) = \int_{-\infty}^{\infty} f(\eta)\rd \eta\,, \quad f(\eta) = \exp \Bigg[ - \frac{1}{2} (r_0+ q) \eta_{\vl}^2 - \sum_{m=2}^{m_0} \frac{u_{2m}}{(2m)!} \eta_{\vl}^{2m}\Bigg].
\ee
Here, $m_0=2$ is the approximation correspondent to the $\rho^4$ model. However, any particular difficulties would arise if one uses higher-order approximations such as: $\rho^6$, which corresponds to $m_0=3$ etc. Keeping with these higher-order approximations, one should use in \eqref{1d1fa} the coefficients $a_{2m}$ where $m>2$.
The coefficients $P_{2m}$ are derived in the way it was done by~\cite{Yu_1987}
\bea\label{1d19fa}
&&
P_2  =I_2\,, \quad P_4 = s^{-d} \big(-I_4 + 3 I_2^2\big),\non
&&
P_6  = s^{-2d} \big(I_6 - 15 I_4 I_2 + 30 I_2^3\big)
\eea
etc., where
\be\label{1d20fa}
I_{2l} = \frac{\displaystyle{\int_{-\infty}^{\infty}} \eta^{2l} f(\eta) \rd \eta}{\displaystyle{\int_{-\infty}^{\infty}} f(\eta) \rd \eta}.
\ee
 The term $s^{-nd}$ in \eqref{1d19fa} is connected with the necessity of transition from the set of values $\vk\in\cB$ to the set $\vk\in\cB_1$ (see~\cite{Yu_1987}), where $d=3$ is the dimension of space.

In terms of parabolic cylinder functions, we have the following:
\begin{align}
%\bea
\label{1d21fa}
Q(r_0) &= (2\piup)^{1/2} \lp \frac{3}{u_0}\rp^{1/4} \text{e}^{x^2/4} U(0,x)\,,\non
P_2 &= \lp \frac{3}{u_0}\rp^{1/2} U(x)\,,\non
P_4 &= s^{-3} \lp \frac{3}{u_0}\rp \varphi(x).
%\eea
\end{align}
The special functions $U(x)$ and $\varphi(x)$
\[
U(x) = U(1,x) \big/ U(0,x)\,, \quad \varphi(x) = 3U^2 (x) + 2x U(x) - 2
\]
are expressed by the functions of parabolic cylinder $U(a,x)$
\[
U(a,x) = \frac{2}{\Gamma(a+\frac{1}{2})} \text{e}^{-x^2/4} \int^\infty_{0} t^{2a} \exp \bigg(-xt^2 - \frac{1}{2} t^4\bigg) \rd t.
\]
The following expression is the result of integration over the variables $\nu_{\vl}$ in \eqref{1d17fa}
\begin{align}
%\bea
\label{1d22fa}
\Xi &= G_\mu\big[Q(r_0)\big]^{N_v} [Q(P)]^{N_1} j_1 \int (\rd \rho)^{N_1} \exp \Bigg[ a_1^{(1)}\sqrt{N_1}\rho_0- \frac{1}{2} \sli_{\vk\in\cB_1}g_1(k) \rhok\rhomk \Bigg] \non
&
\times
\exp \Bigg[ - \frac{a_4^{(1)}}{4!} \frac{1}{N_1} \sli_{\vk_{i}\in\cB_1} \rho_{\vk_{1}}...\rho_{\vk_{4}} \delta_{\vk_{1}+...+ \vk_{4}}\Bigg],
%\eea
\end{align}
moreover,
\begin{align}
%\bea
\label{1d23fa}
Q(P) &= \int_{-\infty}^\infty \rd \nu g(\nu), \quad g(\nu) = \exp \left[ - \sum_{m=1}^{\infty} (2\piup)^{2m} \frac{P_{2m}}{(2m)!} \nu_{\vl}^{2m}\right], \non
a_1^{(1)}& = s^{d/2} w_0\,.
%\eea
\end{align}
For $g_1(k)$ and $a_4^{(1)}$, we have the formulae:
\begin{align}
%\bea
\label{1d24fa}
g_1(k) &= g_1(0) + 2 b k^2, \non
a_4^{(1)} &= \frac{3}{P_4} \varphi(y)\,,\non
g_1(0) &= \lp \frac{3}{P_4}\rp^{1/2} U(y) - q.
%\eea
\end{align}
$q$ is expressed in \eqref{1d14fa},
\be\label{1d25fa}
y = s^{3/2} U(x) \bigg[ \frac{3}{\varphi(x)}\bigg]^{1/2}, \quad x = (r_0 +q)\lp 3/u_0 \rp^{1/2}.
\ee
The following explicit form of the recurrence relations appears between the coefficients of the exponent in the grand partition function after the first stage of integration. Taking into account the formulae for $P_4$ [see \eqref{1d21fa}], $y$ and $x$ [see \eqref{1d25fa}] one has
\be\label{1d26fa}
g_1(0) = N(x) (r_0 + q)-q\,,
\ee
where
\be\label{1d27fa}
N(x) = \frac{y U(y)}{x U(x)}.
\ee
For $a_4^{(1)}$, one derives
\be\label{1d28fa}
a_4^{(1)} = s^{-3} u_0 E(x)\,,
\ee
where
\be\label{1d29fa}
E(x) = s^{2d} \varphi(y) / \varphi(x).
\ee
Use the notion
\[
w_1 =  s a_1^{(1)}, \quad r_1 = s^2 g_1(0)\,, \quad u_1 = s^4 a_4^{(1)}
\]
and write the expressions for $a_1^{(1)}$ from \eqref{1d23fa}, $g_1(0)$ from \eqref{1d26fa} and $a_4^{(1)}$ from \eqref{1d28fa} in the following form:
\begin{align}
%\bea
\label{1d30fa}
w_1 &= s^{\frac{d+2}{2}} w_0\,,\non
r_1 &= s^2 \big[-q + (r_0+q) N(x)\big],\non
u_1 &= s^{4-d} u_0 E(x).
%\eea
\end{align}
The special functions N(x) and E(x) are defined in the above.

According to~\cite{K_2012,Yu_1987}, after making $n$ stages of integration in $\Xi$, we have
\be\label{1d31fa}
\Xi  = 2^{(N_{n+1}-1)/2}G_\mu \big[Q(r_0)\big]^{N_v} Q_1 ...Q_n \big[Q(P_n)\big]^{N_n+1} \int W_{n+1}(\rho) (\rd \rho)^{N_{n+1}},
\ee
where
\bea\label{1d32fa}
&&
Q(d_n) = (2\piup)^{1/2} \Bigg[ \frac{3}{a_4^{(n)}}\Bigg]^{1/4} \exp\lp \frac{x^2_n}{4}\rp U(0,x_n)\,,\non
&&
Q(P_n) = (2\piup)^{-1/2} \left[ \frac{a_4^{(n)}}{\varphi(x_n)}\right]^{1/4} s^{3/4} \exp\lp \frac{y^2_n}{4}\rp U(0,y_n)\,,
\eea
\begin{equation*}
 W_{n_p+1}(\rho) = \exp \Bigg[ a_1^{(n+1)} \sqrt{N_{n+1}} \rho_0 - \frac{1}{2} \sli_{\vk\in\cB_{n+1}} g_{n+1} (k) \rhok\rhomk - \frac{a_4^{(n+1)}}{4!} \frac{1}{N_{n+1}} \sli_{\vk_{i}\in\cB_{n+1}} \rho_{\vk_1}...\rho_{\vk_4} \delta_{\vk_1 + ...+ \vk_4}\Bigg],
\end{equation*}
and also
\be\label{1d33fa}
Q_n = \big[ Q(P_{n-1}) Q(d_n)\big]^{N_n}.
\ee
Moreover,\footnote{Note that $x_n=g_n(B_{n+1},B_n)\bigg[ \frac{3}{a_4^{(n)}}\bigg]^{\frac{1}{2}} $, where
$g_n(B_{n+1}, B_n)=g_n(0)+ q s^{-2n}=s^{-2n}(r_n+q)$.}
\be\label{1d34fa}
x_n = (r_n+q)(3/u_n)^{1/2}, \quad y_n = s^{3/2} U(x_n) \big[3/\varphi(x_n)\big]^{1/2}.
\ee
If $a_1^{(n+1)}=s^{-(n+1)} w_{n+1}$, $g_{n+1}(0) = s^{-2(n+1)}r_{n+1}$, $a_4^{(n+1)}=s^{-4(n+1)}u_{n+1}$, the recurrence relations can be represented as follows:
\begin{align}
%\bea
\label{1d35fa}
w_{n+1} &= s^{\frac{d+2}{2}} w_n\,,\non
r_{n+1} &= s^2 \big[-q + (r_n+q) N(x_n)\big]\,,\non
u_{n+1} &= s^{4-d} u_n E(x_n)
%\eea
\end{align}
with the initial condition \eqref{1d11fa}.
Using the following conditions, one finds the coordinates of the fixed point $w^*, r^*, u^*$
\[
w_n = w_{n+1} = w^*,\quad r_n = r_{n+1} = r^*, \quad u_n = u_{n+1} = u^*.
\]
For $w^* $, there is $w^*=0$, since $s>1$. The equation for $u_{n+1}$ yields
\be\label{1d36fa}
s E(x^*) = 1,
\ee
which juxtaposes own $x^*$ to each $s$. The value $s^*=3.5977$ falls in with $x^*=0$~\cite{K_2012}. From the second equation, we have \eqref{1d35fa}
\be\label{1d37fa}
(u^*)^{1/2}  = q\big(1-s^{-2}\big) \sqrt 3 U(x^*) \big/ \big[y^* U(y^*)\big].
\ee
Therefore, the coordinates of the fixed point of the recurrence relations \eqref{1d35fa} are $w^*=0$, $r^*=-q$, however, $u^*$ should be determined from \eqref{1d37fa}.
Note that the values of $y_n$ from \eqref{1d34fa} are large. Taking this into account, for \eqref{1d35fa} and \eqref{1d37fa} one obtains  the following:
\[
w_{n+1} = s^{\frac{d+2}{2}} w_n\,,
\]
\[
r_{n+1} = s^2 \left[ - q + \frac{\sqrt{u_n}}{\sqrt{3}} \frac{1}{U(x_n)} - \frac{1}{2s^3} \frac{\sqrt{u_n}}{\sqrt{3}} \frac{\varphi(x_n)}{U^3(x_n)}\right],
\]
\[
u_{n+1} = s u_n \frac{\varphi(x_n)}{3 U^4(x_n)} \left[ 1 - \frac{7}{2} s^{-3} \frac{\varphi(x_n)}{U^2(x_n)}\right]
\]
and
\[
(u^*)^{1/2} = q(1-s^{-2}) \sqrt{3} U(x^*) \left[ 1 + \frac{3}{2} (y^*)^{-2} \right].
\]

 The quantity $q$ from \eqref{1d14fa} fails to be a function of temperature, so $r^*$ and $u^*$ fail to be functions of temperature as well. They depend on $s$ and $\alpha_R=\alpha/R_0$. Besides, the formula \eqref{1d8fa} contains the function $\big[U(0)-\Psi(0)/4\big] \big/ \big[U(0)+\Psi(0)(\chi-1)\big]$, which turns into unity in case of $\chi=3/4$.

Using eigenvalues of the transformation matrix $\cal R$
\be\label{1d38fa}
\left(
\begin{array}{lll}
w_{n+1}-w^* \\
r_{n+1}-r^*\\
u_{n+1}-u^*
\end{array}
\right) = \cal R
\left(
\begin{array}{lll}
w_{n}-w^* \\
r_{n}-r^*\\
u_{n}-u^*
\end{array}
\right),
\ee
which, in case of $s=s^*$, are equal to
\be\label{1d39fa}
E_1 = s^{\frac{d+2}{2}}=24.551, \quad E_2 = 8.308, \quad E_3 = 0.374,
\ee
and also using eigenvectors of $\cal R$, one has
\begin{align}
%\bea
\label{1d40fa}
w_n &= w_0 E_1^n,\non
r_n &= r^* + c_1 E_2^n + c_2 R E_3^n,\non
u_n &= u^* + c_1 R_1 E_2^n + c_2  E_3^n,
%\eea
\end{align}
 where $R = R^{(0)}/(u^*)^{1/2}$, $R_1 = R_1^{(0)}(u^*)^{1/2}$, moreover, $R^{(0)}=-0.530$, $R_1^{(0)}=0.162$. The expressions derived in~\cite{K_2012} are valid for the coefficients $c_1$ and $c_2$
\bea\label{1d41fa}
&&
c_1 = \big[r_0 - r^* + (u^* - u_0) R\big] D^{-1}, \non
&&
c_2 = \big[u_0 - u^* + (r^* - r_0) R_1\big] D^{-1},
\eea
and $D=(E_2-E_3)/(R_{22}-E_3)\approx 1.086$. For $R$ both with $R_1$ we have
\begin{align}
%\bea
\label{1d42fa}
R &= R^{(0)} \big/ \sqrt{u^*}, \quad R^{(0)} = \frac{R_{23}^{(0)}}{E_3-R_{22}},\non
R_1 &= R_1^{(0)} \sqrt{u^*}, \quad R_1^{(0)} = \frac{E_2-R_{22}}{R_{23}^{(0)}}.
%\eea
\end{align}
The equation for $T_\text{c}$
\be\label{1d43fa}
c_1(T_\text{c}) = 0
\ee
is of the following form:
\[
1 - \tilde a_2\beta_\text{c} W(0) - r^* - R\big\{a_4\big[\beta W(0)\big]^2 - u^*\big\} = 0.
\]
Since $r^* = -q$, we obtain the equation
\be\label{1d44fa}
1 + q + R^{(0)} \sqrt{u^*} - \tilde a_2 \beta_\text{c} W(0) - R^{(0)} \frac{a_4}{\sqrt{u^*}} \big[\beta_\text{c} W(0)\big]^2 = 0\,,
\ee
where $q$ is defined in \eqref{1d14fa}, $\tilde a_2$ and $a_4$ are the coefficients of the expression \eqref{1d1fa}.

This equation allows us to find the critical temperature of the fluid model as a function of microscopic parameters of the interaction potential and coordinates of the fixed point of the recurrence relations. The physical condition which yields this equation is the existence of critical regime of fluctuations (see~\cite{Yu_1987}) near the critical point. The quantity $\beta_\text{c} W(0) = 4.412$ is the solution of the equation \eqref{1d44fa} for the values in \eqref{1d5fa} and coordinates of the fixed point which are obtained above.

Let us represent $c_1(T)$ and $c_2(T)$ from \eqref{1d41fa} as an expansion in powers of $\tau=(T-T_\text{c})/T_\text{c}$, which is used in section~\ref{sec3} to derive one of the terms of the thermodynamic potential of the model. Applying the equality \eqref{1d11fa} for the expressions of $r_0$ and $u_0$ and taking into account that the coordinates of the fixed point of the recurrence relations \eqref{1d35fa} are not functions of temperature, one has the following:
\bea\label{1d45fa}
&&
c_1 = c_{10} + c_{11}\tau + c_{12}\tau^2,\non
&&
c_2 = c_{20} + c_{21}\tau + c_{22}\tau^2.
\eea
Here, $c_{10}=0$, because of the equation \eqref{1d44fa}, which, actually, is used to determine the critical temperature. For other coefficients, we have the following:
\bea\label{1d46fa}
&&
c_{11} = \beta_\text{c} W(0) D^{-1} \Big[ \tilde a_2 + 2R^{(0)} \beta_\text{c} W(0) a_4 (u^*)^{-1/2}\Big], \non
&&
c_{12} = - \beta_\text{c} W(0) D^{-1} \Big[ \tilde a_2 + 3R^{(0)} \beta_\text{c} W(0) a_4 (u^*)^{-1/2}\Big].
\eea
For the coefficients $c_{2l} (l=0,1,2)$, we get:
\bea\label{1d47fa}
&&
c_{20} = D^{-1} \Big\{ - u^* - R_1^{(0)} \sqrt{u^*} (1+q) + R_1^{(0)} \sqrt{u^*} \tilde a_2 \beta_\text{c} W(0) + a_4 \big[\beta_\text{c} W(0)\big]^2
\Big\}, \non
&&
c_{21} = - D^{-1} \Big\{ R_1^{(0)} \sqrt{u^*} \tilde a_2 \beta_\text{c} W(0) + 2a_4 \big[\beta_\text{c} W(0)\big]^2 \Big\}, \non
&&
c_{22} =  D^{-1} \Big\{ R_1^{(0)} \sqrt{u^*} \tilde a_2 \beta_\text{c} W(0) + 3a_4 \big[\beta_\text{c} W(0)\big]^2 \Big\}.
\eea

\section{A thermodynamic potential of the model}
\label{sec3}

The later calculation of \eqref{1d31fa} is based on the method proposed in~\cite{K_2012}. In case of $T>T_\text{c}$, the thermodynamic potential
\be\label{2d1fa}
\Omega = - k_\text{B} T \ln \Xi
\ee
is represented as follows:
\be\label{2d2fa}
\Omega = \Omega_\mu + \Omega_r + \Omega_\text{CR}^{(+)} + \Omega_\text{LGR}\,,
\ee
where
\bea\label{2d3fa}
&&
\Omega_\mu = - k_\text{B} T \ln G_\mu = - k T \Big\{ \ln g_W + \frac{1}{2} N_v \ln \big[\beta W(0)\big] + N_v (E_\mu - a_0)\Big\},\non
&&
\Omega_r = - k_\text{B} T \ln Q(r_0) = - k T N_v \bigg[ \frac{1}{2} \ln (2 \piup) + \frac{1}{4} \ln \lp \frac{3}{u_0}\rp + \frac{x^2}{4} + \ln U(0,x)\bigg].
\eea
In the critical region of fluctuations, one has
\be\label{2d4fa}
\Omega_\text{CR}^{(+)} = - k_\text{B} T N_v \big( \gamma_{01}+\gamma_{02}\tau +\gamma_{03}\tau^2\big) + \Omega_\text{CR}^{(s)}\,.
\ee
The coefficients $\gamma_{0l}$ meet the expressions
\bea\label{2d5fa}
&&
\gamma_{01} = s^{-3} \frac{f_\text{CR}^{(0)}}{(1-s^{-3})}\,,\non
&&
\gamma_{02} = s^{-3} \frac{c_{11}d_1 E_2}{1-E_2 s^{-3}}\,,\non
&&
\gamma_{03} = s^{-3} \bigg( \frac{c_{12}d_1 E_2}{1-E_2 s^{-3}} + \frac{c^2_{11} d_3 E_2^2}{1-E_2^2 s^{-3}} \bigg),
\eea
which coincide with similar relations for the Ising-like system in an external field~\cite{K_2012}. The quantities $f_\text{CR}^{(0)}$, $d_l$~\cite{K_2012} are expressed by the coordinates of the fixed point.

The term
\be\label{2d6fa}
\Omega_\text{CR}^{(s)} = k_\text{B} T N_v \bar\gamma^{(+)} s^{-3(n_p+1)}
\ee
contains a nonanalytic function of temperature $\tau$ and chemical potential $\mu$. Here,
\be\label{2d7fa}
\bar\gamma^{(+)} = \bar\gamma_1 + \bar\gamma_2 H_\text{c} + \bar\gamma_3 H_\text{c}^2,
\ee
where $\bar\gamma_l$ are constants~\cite{K_2012}
\bea\label{2d8fa}
&&
\bar\gamma_1  = f_\text{CR}^{(0)} \big(1-s^{-3}\big)^{-1},\non
&&
\bar\gamma_2  = d_1 q \big(1-E_2 s^{-3}\big)^{-1},\non
&&
\bar\gamma_3  = d_3 q^2 \big(1-E_2^2 s^{-3}\big)^{-1},
\eea
which have the following values at $s=s^*$
\be\label{2d9fa}
\bar\gamma_1  = 1.529, \quad \bar\gamma_2  = -0.635, \quad \bar\gamma_3  = -0.058.
\ee
For $n_p$, $H_\text{c}$ and $s^{-3(n_p+1)}$, we have
\bea\label{2d10fa}
&&
n_p = - \frac{\ln\big(\tilde h^2 + h_\text{c}^2\big)}{2\ln E_1} - 1\,, \non
&&
H_\text{c} = \tilde \tau \big(\tilde h^2 + h_\text{c}^2\big)^{-1/(2 p_0)},\non
&&
s^{-3(n_p+1)} = \big(\tilde h^2 + h_\text{c}^2\big)^{3/5}.
\eea
Here, $\tilde\tau = \tau(c_{11}/q)$ is the renormalized relative temperature,
\bea\label{2d11fa}
&&
\tilde h = M\big[\beta W(0)\big]^{1/2},\non
&&
h_\text{c} = \tilde\tau^{p_0}.
\eea
Express the exponent
\be\label{2d12fa}
p_0 = \frac{\ln E_1}{\ln E_2}
\ee
in the following form
\be\label{2d13fa}
p_0 = \nu / \mu.
\ee
Here, $\nu$ is the exponent of the correlation length $\xi = \xi^{\pm}|\tau|^{-\nu}$ at $M=0$. It is represented by the formula
\be\label{2d14fa}
\nu = \frac{\ln s^*}{\ln E_2}.
\ee
The behavior of the correlation length at $T=T_\text{c}$ $\big[\xi=\xi^{(c)}M^{-\mu}\big]$ is described by the critical exponent $\mu$
\be\label{2d15fa}
\mu = \frac{2}{d+2}.
\ee
The values $\nu=0.605$ and $p_0=1.512$ are derived for the model $\rho^4$ at $s=s^*$.

Note that the quantities $\Omega_\mu$ and $\Omega_r$ are contained in the expression of the thermodynamic potential $\Omega$ [see \eqref{2d2fa} and \eqref{2d3fa}]. A temperature dependence can be singled out from these terms, but this would cause a renormalization of coefficients of the analytic part of the thermodynamic potential. Of particular interest are the terms of the thermodynamic potential which are the nonanalytic functions of both temperature and chemical potential.

 The part of the thermodynamic potential $\Omega_\text{LGR}$ from \eqref{2d2fa} can be expressed as a sum of two terms~\cite{ypk102}
\be\label{2d16fa}
\Omega_\text{LGR} = \Omega_\text{TR}^{(+)} + \Omega'.
\ee
 The former term $\Omega_\text{TR}^{(+)}$ is a contribution to the thermodynamic potential of the transitional region of fluctuations (from non-Gaussian to Gaussian fluctuations of the order parameter). The latter term $\Omega'$ can be derived using the Gaussian distribution of fluctuations.

The term $\Omega_\text{TR}^{(+)}$ holds for the formula (see~\cite{K_2009})
\be\label{2d17fa}
\Omega_\text{TR}^{(+)} = - k_\text{B} T N_v f_{n_p+1} s^{-3(n_p+1)}.
\ee
The expression for the coefficient $f_{n_p+1}$ is
\be\label{2d18fa}
f_{n_p+1} = \frac{1}{2} \ln y_{n_p} + \frac{9}{4} y^{-2}_{n_p} + \frac{1}{4} x^2_{n_p+1} + \ln U(0, x_{n_p+1}).
\ee
It is easy to derive $x_{n_p+1}$ from the relation
\be\label{2d19fa}
x_{n_p+m} = \bar x E_2^{m-1} H_\text{c} \big(1 + \Phi_q E_2^{m-1} H_\text{c}\big)^{-1/2},
\ee
in case of $m=1$. Here,
\[
\bar x = q(u^*)^{-1/2} \sqrt 3\,, \quad \Phi_q = q(u^*)^{-1/2} R_1^{(0)}.
\]
$H_\text{c}$ is defined in \eqref{2d10fa}, and $y_{n_p}$ should be derived from formula for $y_n$ \eqref{1d34fa} in case of $n=n_p$.

In order to calculate the term $\Omega'$ [see \eqref{2d16fa}], we have to reckon the correspondent expression of the grand partition function $\Xi'$
\be\label{2d20fa}
\Xi' = 2^{(N_{n_p+2}-1)/2} \big[Q(P_{n_p+1})\big]^{N_{n_p+2}} \Xi_{n_p+2}\,,
\ee
where
\begin{align}
%\bea
\label{2d21fa}
\Xi_{n_p+2} &= \int (\rd \rho)^{N_{n_p+2}} \exp\Bigg[ \tilde h \sqrt{N_v} \rho_0 - \frac{1}{2} \sli_{\vk\in\cB_{n_p+2}} g_{n_p+2}(k) \rhok\rhomk  \non
&
- \frac{a_4^{(n_p+2)}}{4!} N_{n_p+2}^{-1} \sli_{{\vk_1,\ldots,\vk_4}\atop{\vk_i\in\cB_{n_p+2}}}\rho_{\vk_1}\ldots\rho_{\vk_4} \delta_{\vk_1+\ldots+\vk_4}\Bigg].
%\eea
\end{align}
The expressions for the coefficients $g_{n_p+2}(k)$ and $a_4^{(n_p+2)}$ are as follows:
\begin{align}
%\bea
\label{2d22fa}
g_{n_p+2} (k) &= g_{n_p+2} (0) + 2 b k^2,\non
g_{n_p+2} (0) &= s^{-2(n_p+2)} r_{n_p+2}\,,\non
a_4^{(n_p+2)} &= s^{-4(n_p+2)} u_{n_p+2}\,,
%\eea
\end{align}
both $r_{n_p+2}$ and $u_{n_p+2}$ meet the expressions
\begin{align}
%\bea
\label{2d23fa}
r_{n_p+2} &= q (-1 + E_2 H_\text{c})\,,\non
u_{n_p+2} &= u^* (1 + \Phi_q E_2 H_\text{c}).
%\eea
\end{align}

\begin{figure}[htbp]
\centering \includegraphics[width=0.47\textwidth]{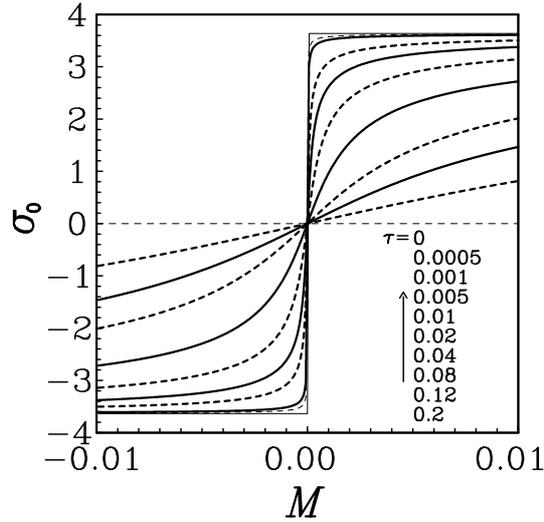}
\caption{The behavior of $\sigma_0$ as a function of the chemical potential $M$ for various values of the relative temperature $\tau$.
The arrow points the correspondence between $\tau$ and curves of $\sigma_0$
at the transition from down to up in the first quadrant of the coordinate plane.}
\label{fig_1fa}
\end{figure}

 The integral in \eqref{2d21fa} is convergent since the coefficient $u_{n_p+2}$ is positive in case of any values of temperature and chemical potential. The coefficient $r_{n_p+2}$ is positive and much larger than $u_{n_p+2}$ in case of small values of the chemical potential ($h_\text{c}\gg \tilde h$). However, in case of large chemical potential $(h_\text{c}\ll\tilde h)$, the coefficient $r_{n_p+2}$ is small and turns into negative if temperature $\tau$ continues to decrease.  It is possible to calculate the expression \eqref{2d21fa} by using the Gaussian distribution and applying the following substitution of variables:
\be\label{2d24fa}
\rhok = \etak + \sqrt{N_v} \sigma_+ \delta_{\vk}\,.
\ee
As a result,
\begin{align}
%\bea
\label{2d25fa}
\Xi_{n_p + 2} &= \text{e}^{N_v E_0(\sigma_+)} \int (\rd \eta)^{N_{n_p+2}} \exp \Bigg[ A_0 \sqrt{N_v} \eta_0 - \frac{1}{2} \sli_{\vk\in\cB_{n_p+2}} \bar g(k) \eta_{\vk} \eta_{-\vk}  \non
&
- \frac{\bar b}{6} N_{n_p+2}^{-1/2} \sli_{{\vk_{1},\ldots,\vk_{3}}\atop{\vk_i \in\cB_{n_p+2}}} \eta_{\vk_{1}}\ldots\eta_{\vk_{3}}
\delta_{\vk_{1} + \ldots+\vk_{3}}
- \frac{\bar a_4}{24} N_{n_p+2}^{-1} \sli_{{\vk_{1},\ldots,\vk_{4}}\atop{\vk_{i} \in\cB_{n_p+2}}} \eta_{\vk_{1}}\ldots\eta_{\vk_{4}}
\delta_{\vk_{1} + \ldots+\vk_{4}} \Bigg].
%\eea
\end{align}
Here,
\begin{align}
%\bea
\label{2d26fa}
E_0(\sigma_+) &= \tilde h \sigma_+ - \frac{r_{n_p+2}}{2} s^{-2(n_p+2)} \sigma_+^2 - \frac{u_{n_p+2}}{24} s^{-(n_p+2)} \sigma_+^4\,,\non
A_0 &= \tilde h - r_{n_p+2} s^{-2(n_p+2)} \sigma_+ - \frac{u_{n_p+2}}{6} s^{-(n_p+2)} \sigma_+^3\,,\non
\bar g(k) &= \bar g(0) + 2 b k^2,\non
\bar g(0) &= r_{n_p+2} s^{-2(n_p+2)} + \frac{u_{n_p+2}}{2} s^{-(n_p+2)}\sigma_+^2\,,\non
\bar b &= u_{n_p+2} s^{-5(n_p+2)/2} \sigma_+\,, \quad
\bar a_4 = u_{n_p+2} s^{-4(n_p+2)}.
%\eea
\end{align}
As in~\cite{K_2009}, the magnitude of the shift $\sigma_+$ is found from the condition
\be\label{2d27fa}
\frac{\partial E_0(\sigma_+)}{\partial \sigma_+} = 0.
\ee
Recall the expression of $A_0$ from \eqref{2d26fa} to write the equation
\be\label{2d28fa}
A_0 = 0\,,
\ee
with a solution chosen in the form of
\be\label{2d29fa}
\sigma_+ = \sigma_0 s^{-(n_p+2)/2}.
\ee
The solutions of the gotten cubic equation for $\sigma_0$
\be\label{2d30fa}
\sigma_0^3 + p \sigma_0 + q = 0\,,
\ee
where
\[
p = 6 \frac{r_{n_p+2}}{u_{n_p+2}}\,, \quad q = - 6 \frac{s^{5/2}}{u_{n_p+2}} \frac{\tilde h}{\big(\tilde h^2 + h_\text{c}^2\big)^{1/2}}\,,
\]
is analyzed by~\cite{K_2012,K_2009}. In general case $(T\neq T_\text{c}, M\neq 0)$, the solution of the equation \eqref{2d30fa} is a function of both chemical potential and temperature. For all $\tau>0$, $M\neq 0$ the real root of this equation is found using the Cardano solution $\sigma_0 = A+B$ (see~\cite{Yu_1987,Korn_1974}), where
\[
A = \big(-q/2 + Q^{1/2}\big)^{1/3}, \quad B = \big(-q / 2 - Q^{1/2}\big)^{1/3},
\]
\[
Q = (p/3)^3 + (q/2)^2.
\]
Direct calculation shows that $Q$ is positive in case of $T>T_\text{c}$.
The plot of real solution $\sigma_0$ as a function of the chemical potential $M$ for various values of $\tau$ is shown in figure~\ref{fig_1fa}.

The Gaussian distribution of fluctuations is a basis (a zero-order approximation at $k\neq 0$) for integration in \eqref{2d25fa} over the variables $\etak$ with $k\neq 0$. Singling out in \eqref{2d25fa} the terms with $k = 0$ and integrating over $\etak$ with $k\neq 0$, we obtain
\be\label{2d31fa}
\Xi_{n_p+2} = \text{e}^{N_vE_0(\sigma_+)} \prod_{{\vk\neq 0}\atop{\vk\in\cB_{n_p+2}}} \big[ \piup / \bar g(k) \big]^{1/2} \Xi^{(0)}_{n_p+2}\,.
\ee
Here,
\be\label{2d32fa}
\Xi^{(0)}_{n_p+2} = \int_{-\infty}^{+\infty} \rd \eta_0 \exp \bigg[ A_0 \sqrt{N_v} \eta_0 - \frac{1}{2} \bar g(0) \eta_0^2 - \frac{\bar b}{6} N_{n_p+2}^{-1/2} \eta_0^3 - \frac{\bar a_4}{24} N^{-1}_{n_p+2} \eta_0^4 \bigg].
\ee
The next stage of calculations is a return to the variable $\rho_0$ using the ``reverse'' to \eqref{2d24fa} change of variables
\be\label{2d33fa}
\eta_0 = \rho_0 - \sqrt{N_v} \sigma_+
\ee
 in \eqref{2d32fa}. The following expression is the result of integration over $\rho_0$ in the term \eqref{2d31fa} of the total thermodynamic potential using the Laplace method
\bea\label{2d34fa}
\Omega_{n_p+2} = - k_\text{B} T N_v E_0(\sigma_+) - \frac{1}{2} k_\text{B} T N_{n_p+2} \ln \piup + \frac{1}{2} k_\text{B} T \sli_{{\vk \neq 0}\atop{\vk\in\cB_{n_p+2}}} \ln \bar g(k).
\eea
Here,
\be\label{2d35fa}
E_0(\sigma_+) = e_0^{(+)} \tilde h s^{-(n_p+1)/2} - e_2^{(+)} s^{-3(n_p+1)},
\ee
where the following notation is used
\be\label{2d36fa}
e_0^{(+)} = \sigma_0 s^{-1/2}, \quad e_2^{(+)} = \frac{\sigma_0^2}{2} s^{-3} \bigg( r_{n_p+2} + \frac{u_{n_p+2}}{12} \sigma_0^2 \bigg).
\ee
Note that both the change of variables \eqref{2d33fa} and the substitution $\rho_0=\sqrt{N_v}\rho$ in \eqref{2d32fa} cause an appearance of a sharp maximum of the integrand expression at the point $\bar\rho$. Both the extremum condition of the integrand, from which it is possible to define $\bar\rho$, and the representation $\bar\rho=\bar\rho' s^{-(n_p+2)/2}$ lead to the same equation (with the same coefficients) as \eqref{2d30fa}, where the role of $\sigma_0$ is played by $\bar\rho'$. The expressions of $\bar\rho'$ and $\sigma_0$ coincide. The quantities of $\bar\rho$ and $\sigma_+$ \eqref{2d29fa} are also similar. Thus, the integrand $E_0(\bar\rho)$ at the point $\bar\rho$ is represented as $E_0(\sigma_+)$ [see \eqref{2d35fa}] with coefficients \eqref{2d36fa}, where $\bar\rho'$ is changed to $\sigma_0$. The sum over $\vk\in\cB_{n_p+2}$ in \eqref{2d34fa} is calculated using the transition to the spherical Brillouin zone. Integration over $k$:
\begin{align}
%\bea
\label{2d37fa}
\frac{1}{2} \sli_{\vk\in\cB_{n_p+2}} \ln \bar g(k) &= N_{n_p+2} \left[ \frac{1}{2} \ln\big(1+a^2\big) - (n_p+2)\ln s + \frac{1}{2} \ln r_R -\frac{1}{3} + \frac{1}{a^2} - \frac{1}{a^3} \arctan{a} \right],\non
&
a = \frac{\piup}{c} \lp \frac{2b}{r_R}\rp^{1/2}, \quad r_R = r_{n_p+2} + \frac{u_{n_p+2}}{2} \sigma_0^2.
%\eea
\end{align}

The expressions derived for  $Q(P_{n_p+1})$ [see \eqref{1d32fa}], $\Omega_{n_p+2}$ \eqref{2d34fa} and $\frac{1}{2} \sli_{\vk\in\cB_{n_p+2}}\ln \bar g(k)$ \eqref{2d37fa} are used to write the part of the thermodynamic potential, which is correspondent to  \eqref{2d20fa}, as a sum of two terms
\be\label{2d38fa}
\Omega' = \Omega_0^{(+)} + \Omega'_G\,.
\ee
The term
\be\label{2d39fa}
\Omega_0^{(+)} = - k_\text{B} T N_v E_0(\sigma_+)
\ee
is a part of the thermodynamic potential connected to the variable $\rho_0$. For $\Omega'_G$, one has
\be\label{2d40fa}
\Omega'_G = - k_\text{B} T N_{n_p+2} f_G\,.
\ee
The coefficient $f_G$ is as follows:
\be\label{2d41fa}
f_G = \bigg[ - \frac{1}{2} \ln 3 + 2 \ln s + \frac{1}{2} \ln u_{n_p+1} - \ln r_R - \ln U(x_{n_p+1}) - \frac{3}{4} y^{-2}_{n_p+1}  - f''_G \bigg] \Big/ 2.
\ee
Here,
\bea\label{2d42fa}
&&
u_{n_p+1} = u^* (1 + \Phi_q H_\text{c})\,,\non
&&
x_{n_p+1} = \bar x H_\text{c} (1 + \Phi_q H_\text{c})^{-1/2}.
\eea
$r_R$ and $y_{n_p+1}$ are defined in \eqref{2d37fa} and \eqref{1d34fa}, respectively, and for $f''_G$, one gets
\be\label{2d43fa}
f''_G = \ln\big(1 + a^2\big) - \frac{2}{3} + \frac{2}{a^2} - \frac{2}{a^3} \arctan{a}.
\ee

The part of the thermodynamic potential can be calculated  $\Omega_\text{LGR}$ \eqref{2d16fa} using the expressions of both $\Omega_\text{TR}^{(+)}$ \eqref{2d17fa} and $\Omega'$ \eqref{2d38fa}.

The complete expression of the thermodynamic potential of a cell fluid model is obtained based on~\eqref{2d2fa} in the way of gathering derived terms from all of the regimes of fluctuations for temperature $T>T_\text{c}$. The quantity $\ln g_W$ [see \eqref{1d2fa}] contained in $\Omega_\mu$ [see \eqref{2d3fa}] is found as a result of transition to the spherical Brillouin zone and integration over $k$:
\begin{align}\label{2d44fa}
\ln g_W &= N_v f_W\,, \non
f_W &= - \frac{1}{2} \ln (2\piup) - \frac{1}{2}\ln \big[\beta W(0)\big] + f'_{W}\,,  \non
f'_W &= - \frac{1}{2} \ln \big[1-(a')^2\big] + \frac{1}{3} + \frac{1}{(a')^2} - \frac{1}{2(a')^3} \ln{ \bigg| \frac{1+a'}{1-a'}\bigg|}\,, \non
a' &= \frac{\piup}{c} (2b)^{1/2}.
\end{align}
The complete expression of the thermodynamic potential, which is equivalent to \eqref{2d2fa}, is represented in the form of three terms
\be\label{2d45fa}
\Omega = \Omega_a + \Omega_s^{(+)} + \Omega_0^{(+)}.
\ee
The analytic part of the thermodynamic potential $\Omega_a$ is as follows:
\be\label{2d46fa}
\Omega_a = - k_\text{B} T N_v \big( \gamma_{01} + \gamma_{02} \tau + \gamma_{03} \tau^2 \big) + \Omega_{01}\,,
\ee
where
\begin{align}\label{2d47fa}
\Omega_{01} &= - k_\text{B} T N_v \left(  E_\mu + \gamma_a \right), \\
\gamma_a &= f'_{W} - a_0 + \frac{1}{4} \ln \lp \frac{3}{u_0} \rp + \frac{x^2}{4} + \ln U(0, x).\nonumber
\end{align}
The quantities $E_\mu$, $u_0$, $x$ and $a'$ are defined in \eqref{1d2fa}, \eqref{1d11fa}, \eqref{1d25fa} and \eqref{2d44fa}.
The term $\Omega_s^{(+)}$ is a sum of nonanalytic contribution. It has the following form
\be\label{2d48fa}
\Omega_s^{(+)} = - k_\text{B} T N_v \gamma_s^{(+)} \big(\tilde h^2 + h_\text{c}^2\big)^{\frac{d}{d+2}}.
\ee
Here,
\be\label{2d49fa}
\gamma_s^{(+)} = f_{n_p+1} - \bar\gamma^{(+)} + f_G / s^3.
\ee
The shift of the variable $\rho_0$ is determined by the quantity $\sigma_0$, which is contained in the coefficients $e_0^{(+)}$ and $e_2^{(+)}$ of the term
\bea\label{2d50fa}
&&
\Omega_0^{(+)} = - k_\text{B} T N_v \left[ e_0^{(+)} \tilde h \big(\tilde h^2 + h_\text{c}^2\big)^{\frac{d-2}{2(d+2)}} - e_2^{(+)} \big(\tilde h^2 + h_\text{c}^2\big)^{\frac{d}{d+2}} \right].
\eea
The expressions for the coefficients $e_0^{(+)}$ and $e_2^{(+)}$ are presented in \eqref{2d36fa}.

\section{An equation of state of the model at $T\geqslant T_\text{c}$ with effects of \\ fluctuations taken into account}
\label{sec4}

We derived the thermodynamic potential $\Omega = -k_\text{B} T\ln \Xi$ [see \eqref{2d45fa}] for the cell fluid model taking into account non-Gaussian fluctuations of the order parameter. Using the expression of the logarithm of the grand partition function
\begin{align}\label{3d1fa}
\ln \Xi &= N_v \bigg\{ P_a (T) + E_\mu + \Big[ \gamma_s^{(+)} - e_2^{(+)}\Big] \lp \tilde h^2 + h_\text{c} ^2\rp^{\frac{d}{d+2}} + e_0^{(+)} \tilde h \lp \tilde h^2 + h_\text{c}^2\rp^{\frac{d-2}{2(d+2)}}  \bigg\}, \\
 P_a (T) &= \gamma_a + \gamma_{01} + \gamma_{02}\tau + \gamma_{03} \tau^2  \nonumber
\end{align}
it is possible to get the expression of the pressure $P$ as a function of the temperature $T$ and the chemical potential $\mu$ using the well-known formula
\be\label{3d2fa}
P V = k_\text{B} T \ln \Xi.
\ee
Having the grand partition function, it is possible to calculate the average number of particles
\be\label{3d3fa}
\bar{N} = \frac{\partial \ln\Xi}{\partial \beta\mu}.
\ee
The latter relation is applicable to express the chemical potential via either the number of particles or the relative density
\be\label{3d4fa}
\bar n = \frac{\bar N}{N_v} = \lp \frac{\bar N}{V}\rp v\,,
\ee
 where $v$ is the volume of a cubic cell, which is a parameter of the model in use. Combining the equalities~\eqref{3d2fa} and \eqref{3d3fa}, let us find the pressure $P$ as a function of the temperature $T$ and the relative density $\bar n$, which is the equation of state of the model we study.

Using \eqref{3d1fa}, \eqref{3d3fa} and \eqref{3d4fa}, one obtains
\be\label{3d5fa}
\bar n = \frac{\partial E_\mu}{\partial \beta\mu} + \frac{\partial}{\partial\beta \mu} \left[ \gamma_s^{(+)} \lp \tilde h^2 + h_\text{c}^2 \rp^{\frac{d}{d+2}}\right] + \frac{\partial}{\partial\beta\mu} \left[ e_0^{(+)} \tilde h \lp \tilde h^2 + h_\text{c}^2 \rp^{\frac{d-2}{2(d+2)}}
- e_2^{(+)} \lp \tilde h^2 + h_\text{c}^2 \rp^{\frac{d}{d+2}} \right].
\ee
Based on \eqref{3d5fa} and taking into account \eqref{3d6fa} (see Appendix A) one has
\be\label{3d10fa}
\bar n = - M - \tilde a_1 + \frac{1}{\beta W(0)} a_{34} + \sigma_{00}^{(+)} \big( \tilde h^2 + h_\text{c}^2 \big)^{\frac{d-2}{2(d+2)}}.
\ee
Here, the coefficient
\be\label{3d11fa}
\sigma_{00}^{(+)} = e_0^{(+)} \frac{1}{\big[\beta W(0)\big]^{1/2}} \lp 1 + \frac{d-2}{d+2} \frac{\tilde h^2}{\tilde h^2 + h_\text{c}^2} \rp +
e_{00}^{(+)} \frac{\tilde h}{\big(\tilde h^2 + h_\text{c}^2\big)^{1/2}} + e_{01}^{(+)} \big( \tilde h^2 + h_\text{c}^2 \big)^{1/2},
\ee
beside $e_0^{(+)}$ from \eqref{2d36fa}, contains the quantities
\be\label{3d12fa}
e_{00}^{(+)} = \frac{2d}{d+2} \frac{1}{\big[\beta W(0)\big]^{1/2}} \left[ \gamma_s^{(+)} - e_2^{(+)} \right]
\ee
and
\be\label{3d13fa}
e_{01}^{(+)} = \frac{\partial \gamma_s^{(+)}}{\partial\beta\mu} - \frac{\partial \text{e}^{(+)}_2}{\partial\beta\mu}.
\ee
The final expression for $\sigma_{00}^{(+)}$ is found having calculated the coefficient $e_{01}^{(+)}$ (see Appendix B).

 Taking into account the expressions for $\tilde a_1$ [see \eqref{1d2fa}], $d(0)$ and $\tilde a_2$ [see \eqref{1d3fa}] the equation \eqref{3d10fa} gets the following form
\be\label{3d37fa}
\bar n = n_g - M + \sigma_{00}^{(+)} \big( \tilde h^2 + h_\text{c}^2 \big)^{\frac{d-2}{2(d+2)}}.
\ee
Here,
\be\label{3d38fa}
n_g  =- a_1 - a_2 a_{34} + \frac{a_4}{3} a_{34}^3\,,
\ee
and the coefficient $\sigma_{00}^{(+)}$ is defined in \eqref{3d35fa}. The quantity $\tilde h$ is a function of $M$, see \eqref{2d11fa}. The nonlinear equation \eqref{3d37fa} describes a connection between the density $\bar n$ and the chemical potential $M$. It can be rewritten in the form
\be\label{3d39fa}
\bar n - n_g + M = \left[ \frac{M b_1^{(+)}}{b_2^{(+)}}\right]^{1/5} \sigma_{00}^{(+)}.
\ee
Herefrom we obtain
\be\label{3d40fa}
M b_1^{(+)} = \left[ \frac{\bar n - n_g + M}{\sigma_{00}^{(+)}} \right]^5 b_2^{(+)}
\ee
or
\be\label{3d41fa}
b_3^{(+)} M^{1/5}  = \bar n - n_g + M.
\ee
The coefficients $b_1^{(+)}$, $b_2^{(+)}$ and $b_3^{(+)}$ are given by the formulae
\be\label{3d42fa}
b_1^{(+)} = \big[ \beta W(0) \big]^{1/2}, \quad b_2^{(+)} = \frac{\alpha}{(1+\alpha^2)^{1/2}}\,, \quad
b_3^{(+)} = \left[ \frac{b_1^{(+)}}{b_2^{(+)}} \right]^{1/5} \sigma_{00}^{(+)}.
\ee
Following the equation $M^{1/5}\big(b_3^{(+)} - M^{4/5}\big)=0$ at $\bar n = n_g$, one has such values of the chemical potential~$M$:
\[
M_1 = 0\,, \quad M_{2,3} = \pm \left[ b_3^{(+)} \right]^{5/4}.
\]
The points $M_{m1, m2}$ of the extremum of $\bar n$ are found from the following equation
\[
\frac{1}{5} b_3^{(+)} M_{m1,m2}^{-4/5} - 1 = 0.
\]
Evidently,
\[
M_{m1,m2} = \pm \left[ \frac{b_3^{(+)}}{5} \right]^{5/4}.
\]
Easy to see that there exists an interval of values
\be\label{3d43fa}
M_{m2} < M < M_{m1}\,,
\ee
 where $\bar n$ increases with the growth of $M$. For all $M_{m2}>M>M_{m1}$, a decrease of $\bar n$ is observed with an increase of $M$, which is not a reflection of a physical phenomenon. The interval \eqref{3d43fa} is physical.

\begin{figure}[b!]
\includegraphics[width=0.452\textwidth]{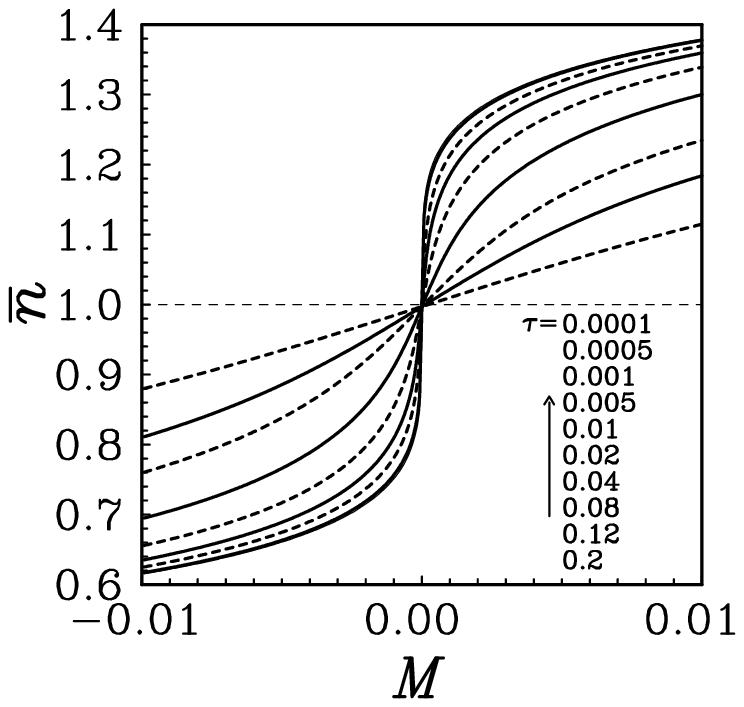}
\hfill
\includegraphics[width=0.47\textwidth]{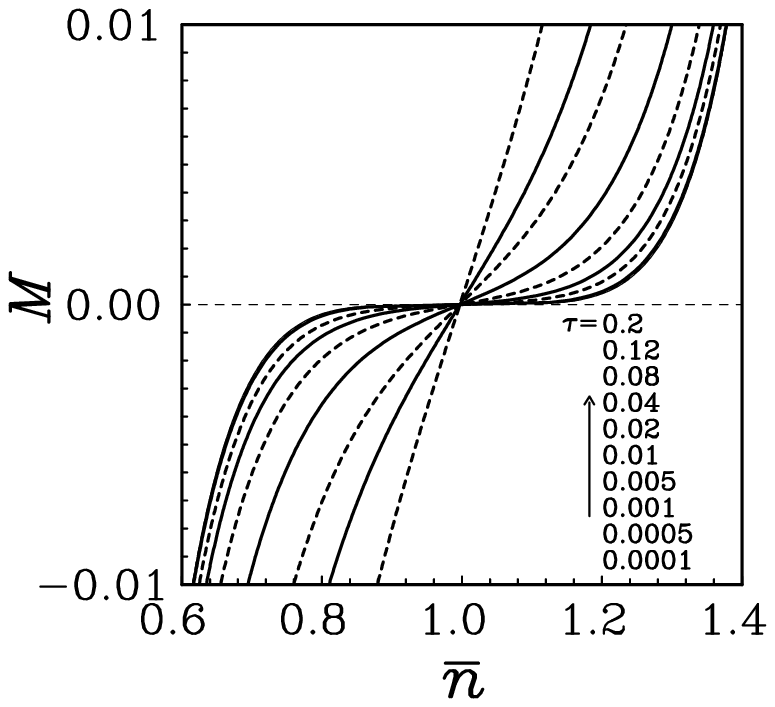}
\\
\parbox[t]{0.452\textwidth}{
\vspace{-0.3cm}
\caption{Plot of the relative density $\bar n$ as a function of the chemical potential $M$.}
\label{fig_2fa}
}
\hfill
\parbox[t]{0.47\textwidth}{
\vspace{-0.3cm}
\caption{Plot of the chemical potential $M$ as a function of the relative density $\bar n$.}
\label{fig_3fa}
}
\end{figure}

Figure~\ref{fig_2fa} and figure~\ref{fig_3fa} show, respectively, the plots of functions $\bar n(M)$ and $M(\bar n)$ which are obtained from the equation \eqref{3d41fa} for different temperatures $\tau$.

\begin{figure}[t!]
	\centering \includegraphics[width=0.54\textwidth]{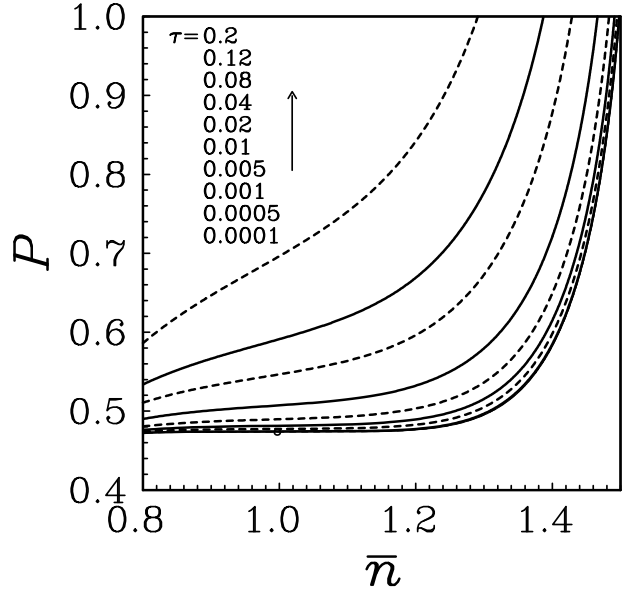}
	\caption{Plot of the pressure $P$ as a function of the relative density $\bar n$ for different temperatures $\tau$. The critical point ($\bar n_\text{c} = 0.997$, $P_\text{c} = 0.474$) is denoted by the symbol $\circ$.}
	\label{fig_4fa}
\end{figure}

Note that the curves for temperatures $\tau = 0.001$, $\tau = 0.0005$, $\tau = 0.0001$ coincide in these figures, as well as in figure~\ref{fig_4fa}.
In case of $M\ll 1$, there is a possibility to express the chemical potential $M$ as a function of the relative density $\bar n$ using the equation \eqref{3d41fa}. The fixed-point iteration method (method of successive approximations) is suitable for solving this equation. In the zero-order approximation, the term $M$ on the right-hand side of the equation \eqref{3d41fa} is neglected and the consequent result is
\be\label{3d44fa}
M^{(0)} = \left[ \frac{\bar n - n_g}{\sigma_{00}^{(+)}}\right]^5 \frac{b_2^{(+)}}{\big[\beta W(0)\big]^{1/2}}\,.
\ee
The outcome of the first-order approximation, assuming that $M=M^{(0)}$ on the right-hand side of the equation \eqref{3d41fa}, is either
\[
M^{(1)} = \left[ \frac{\bar n - n_g + M^{(0)}}{\sigma_{00}^{(+)}}\right]^5 \frac{b_2^{(+)}}{\big[\beta W(0)\big]^{1/2}}
\]
or
\[
M^{(1)} =  M^{(0)} \left[ 1 + \frac{M^{(0)}}{\bar n - n_g}\right]^5.
\]
Henceforth, the zero-order approximation of \eqref{3d44fa} is used for $M$, namely, the case $M=M^{(0)}$ is considered. Note that the region $M>0$ is in agreement with positive values of $\bar n - n_g$, and $\bar n <n_g$ meet~$M<0$. The chemical potential is equal to zero when $\bar n = n_g$. Taking into account \eqref{3d1fa} and \eqref{3d2fa}, at $T>T_\text{c}$ the following equation of state is derived:
\be\label{3d45fa}
\frac{P v}{k_\text{B} T} = P_a (T) + E_\mu + \left[ \gamma_s^{(+)} - e_2^{(+)} \right] \lp \tilde h^2 + h_\text{c}^2 \rp^{\frac{d}{d+2}} + e_0^{(+)} \tilde h \lp \tilde h^2 + h_\text{c}^2 \rp^{\frac{d-2}{2(d+2)}}.
\ee
Here, the quantity $E_\mu$ is expressed in \eqref{1d2fa} where the chemical potential from the formula \eqref{3d44fa} should be substituted for $M$. Using the formula \eqref{3d44fa} in the expression of $\tilde h$ [see \eqref{2d11fa}], one gets
\be\label{3d46fa}
\tilde h = \left[ \frac{\bar n - n_g}{\sigma_{00}^{(+)}} \right]^5 b_2^{(+)}.
\ee
 The definition of $h_\text{c}$ is in \eqref{2d11fa}. The formula \eqref{3d37fa} is used for determining $\tilde h^2 + \tilde h^2_\text{c}$ and then rewriting the equation \eqref{3d45fa} as follows:
\be\label{3d47fa}
\frac{P v}{k_\text{B} T} =  P_a (T) + E_\mu + \left[ \frac{\bar n - n_g}{\sigma_{00}^{(+)}} \right]^6 \Big[ e_0^{(+)} \frac{\alpha}{(1+\alpha^2)^{1/2}} + \gamma_s^{(+)} - e_2^{(+)} \Big].
\ee

In case of $T=T_\text{c}$, it is possible to describe the behavior of the system by the expression \eqref{3d47fa}. In this situation, $\alpha\rightarrow\infty$. Moreover, $H_\text{c}\rightarrow 0$ and $r_{n_p+2}$, $u_{n_p+2}$, $x_{n_p+2}$ fail to be the functions of the chemical potential. The coefficients $\sigma_0$, $e_0^{(+)}$, $e_2^{(+)}$ and $b_2^{(+)}$ become some constants. The equation of state at $T=T_\text{c}(\tau=0)$ is of the following form:
\bea\label{3d48fa}
&&
\frac{P v}{k_\text{B} T_\text{c}} = P_a (T_\text{c}) + E_\mu|_{T_\text{c}} + \frac{5}{6} (\bar n - n_g)^6  \frac{\big[\beta_\text{c} W(0)\big]^{1/2}}{\big[\sigma_{00}^{(+)}(T_\text{c})\big]^5},
\eea
where
\be\label{3d49fa}
\sigma_{00}^{(+)}(T_\text{c}) = \frac{6}{5} \frac{1}{\big[\beta_\text{c} W(0)\big]^{1/2}} \Big[ e_0^{(+)} + \gamma_s^{(+)}-e_2^{(+)}\Big].
\ee
 At the critical point, the density $\bar n_\text{c}$ is expressed by the quantity $n_g$, namely
\begin{equation}\label{3d50fa}
    \bar n_\text{c} = n_g\,.
\end{equation}

The estimated values of parameters of the critical point for sodium are shown in table~\ref{tab1} . These values are obtained in different ways: from our present research for the cell fluid model, from Monte Carlo simulation data for the continuous system with the Morse potential in the grand canonical ensemble~\cite{singh}. In addition table~\ref{tab1} contains the known values from experiments for sodium~\cite{hensel}, which are not related to any interaction potential. Therefrom, one can see that the use of the continuous Morse potential and its lattice counterpart yields results of the same order.
\vspace{-3mm}
\begin{table}[t!]
\caption{ The values of the temperature ($T_\text{c}$), density ($\bar n_\text{c}$) and pressure ($P_\text{c}$) at the critical point from the present research (theory), simulations and experiment. The values are presented in the form of reduced dimensionless units, where temperature $k_\text{B} T = k_\text{B} T'/D$, density $\bar n = \rho R_0^3$ , and pressure $P = P' R_0^3 / D$ \big(the quantities $T'$, $\rho$, $P'$ have dimensionality, for example, $[T'] = [\text K]$, $ [\rho] = [1/\text{m}^3]$, $[P'] = [\text{Pa}]$\big).}
\vspace{2mm}
  \centering
  \begin{tabular}{|l|c|c|c|}
  \hline
   & $k_\text{B} T_\text{c}$ & $\bar n_\text{c}$ & $P_\text{c}$ \\ \hline
  Theory (cell fluid model) & 4.028 & 0.997 & 0.474 \\ \hline
  Simulations (continuous system with the Morse potential) & 5.874 & 1.430 & 2.159 \\ \hline
  Experiment & 3.713 & 1.215 & 0.415 \\
  \hline
\end{tabular}
  \label{tab1}
\end{table}
\vspace{2mm}

 The expressions \eqref{3d47fa} and \eqref{3d48fa} match the case where the term $M$ is neglected on the right-hand side of the equation \eqref{3d41fa}. If this term is not neglected, the numerical results give the behavior of pressure $P$ as a function of an increasing density $\bar n$ for various values of $\tau$ which is shown in figure~\ref{fig_4fa}.

\begin{figure}[b!]
\includegraphics[width=0.52\textwidth]{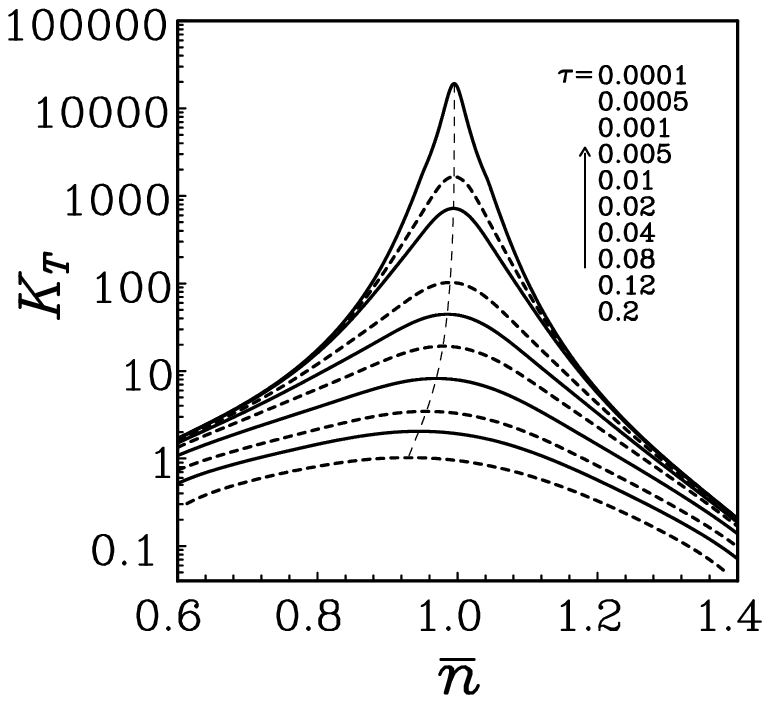}
\hfill
\includegraphics[width=0.5\textwidth]{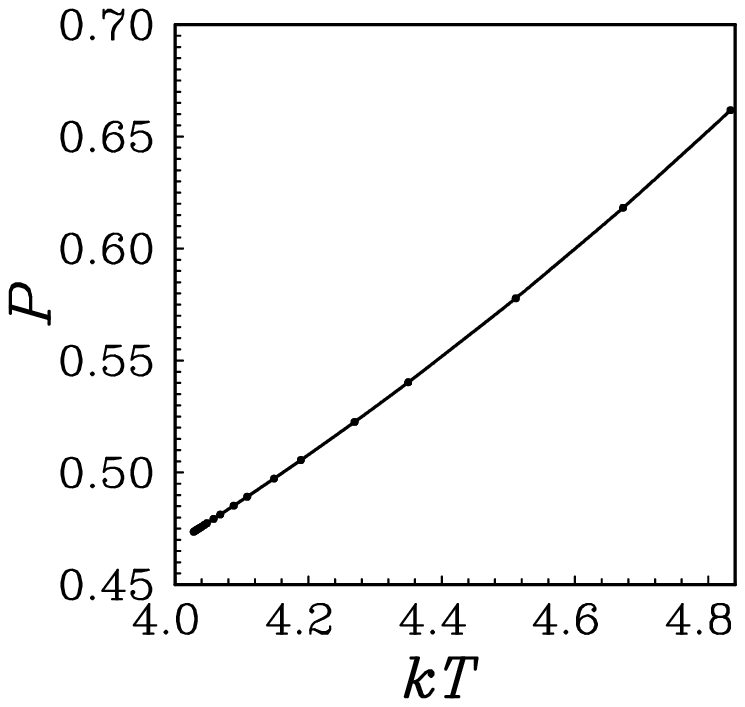}
\\
\parbox[t]{0.52\textwidth}{
\vspace{-0.3cm}
\caption{Plot of the isothermal compressibility $K_T$ as a function of the density
$\bar n$ for different values of the relative temperature $\tau$.}
\label{fig_5fa}
}
\hfill
\parbox[t]{0.5\textwidth}{
\vspace{-0.3cm}
\caption{Plot of the pressure as a function of tem\-pe\-ra\-tu\-re at the extremum points of the compressibility.}
\label{fig_6fa}
}
\end{figure}

The isothermal compressibility $K_T = (\partial \eta/\partial \tilde p)_T/ \eta$, where $\eta = \bar n/\bar n_\text{c}$, $\tilde p = P/P_\text{c}$, is presented in figure~\ref{fig_5fa}.

It is known that discontinuous changes of the properties of the fluid along the first-order binodal that culminates in a critical point can be extended into the supercritical region as the Widom line~\cite{widom,bryk}. This line is defined by the maxima of some thermodynamic quantities such as the constant-pressure specific heat, the correlation length, etc.
Taking into account the extremum values of $K_T$ (see figure~\ref{fig_5fa}) it is possible to plot the Widom-like line of a supercritical cell fluid. The temperature dependence of the pressure $P$ at the extrema points of $K_T$ is shown in figure~\ref{fig_6fa}.

\section{Conclusions}
\label{concl}

An explicit expression for the thermodynamic potential of the cell fluid
model is obtained in the collective variables representation. The basic idea of
the calculation of the thermodynamic potential near $T_\text{c}$ at
a microscopic level lies in the separate inclusion of contributions from
short-wave and long-wave modes of the order parameter oscillations. The short-wave
modes are characterized by the presence of the RG symmetry and are described
by a non-Gaussian measure density. In this case, the RG method is used.
The corresponding RG transformation can be
related to the case of the one-component magnet in the external field~\cite{KPP_2006PA}. The approach which we propose is based on the use of a non-Gaussian density of measure.
The inclusion of short-wave oscillation modes leads to a renormalization of
the dispersion of the distribution describing long-wave modes. The way of taking into account the contribution of long-wave modes of oscillations to the thermodynamic potential of the cell fluid model is qualitatively different from the method of calculating the short-wave part of the grand partition function. The calculation of this contribution grounds on the use of
the Gaussian density of measure as the basis density. The dispersion of this Gaussian distribution becomes a non-analytic function of temperature and density due to consideration of short-range fluctuations. We have developed
a direct method of calculating the thermodynamic potential including both types of oscillation modes in a supercritical region.

The nonlinear equation which links the relative density $\bar n$ and the chemical potential $M$  is derived and investigated. The expressions for
coefficients of this equation are represented as functions of the ratio
of the renormalized chemical potential to the renormalized temperature.
The quantity $\bar n$ corresponding to $M=0$ is obtained. The interval of
values of the chemical potential, within which the density $\bar n$ increases with
the growth of $M$ is indicated. Reduction of $\bar n$ with an increase of $M$
beyond this range does not reflect the physical nature of the phenomenon.
The chemical potential is expressed in terms of the density.

The equation of state for the case of temperatures above the critical value of $T_\text{c}$ obtained in the present research provides the pressure as a function of temperature and density. The equation of state corresponding to the case of $T=T_\text{c}$ is also derived.

The main advantage of this equation of state is the presence of relations connecting its coefficients with the fixed-point coordinates and the microscopic parameters of the interaction potential. It shows the capability of the collective variables method to be efficient in describing both universal and non-universal characteristics of the system as functions of microscopic parameters. This possibility is rather unusual within an RG approach since, for example, it is well-known that the RG of perturbative field theory is incapable of explicitly accounting for a non-universal effect of a particular microscopic parameter of a specific system.

Using the analytic expressions obtained in this study, we made numerical estimations. In particular, the results of the estimates are represented as plots of compressibility and isotherms of pressure as functions of density. It is also shown that the extrema of compressibility form a line on the $(P, T)$ plane, which is similar to the Widom line.

The technique elaborated here for the derivation of the equation of state at temperatures above the $T_\text{c}$ is envisaged to be generalised to the case of $T<T_\text{c}$. The calculations can be extended to the higher non-Gaussian distribution (the $\rho^6$ model)~\cite{ypk102,ypk202}.

\section*{Acknowledgements}

%The authors are thankful to their colleague Ihor Omelyan for valuable advices and discussions.

Authors thank their colleague Ihor Omelyan for his valuable help with numerical calculations.

This work was partly supported by the European Commission under the project STREVCOMS PIRSES-2013-612669.

\newpage
\appendix
\section{ Deriving the expression of the relative density $\bar n$}
%\label{appA}
%\setcounter{equation}{0}

In order to derive the expression of the relative density $\bar n$ \eqref{3d5fa}, one should take the following partial derivatives
\bea\label{3d6fa}
&&
\frac{\partial E_\mu}{\partial \beta\mu} = - M - \tilde a_1 + \frac{a_{34}}{\beta W(0)}\,,\non
&&
\frac{\partial}{\partial \beta\mu} \left[ \gamma_s^{(+)} \lp \tilde h^2 + h_\text{c}^2 \rp^{\frac{d}{d+2}}\right] = \lp \tilde h^2 + h_\text{c}^2 \rp^{\frac{d-2}{2(d+2)}} \left[ \lp \tilde h^2 + h_\text{c}^2 \rp^{1/2} \frac{\partial \gamma_s^{(+)}}{\partial\beta\mu} + \frac{2d}{d+2} \gamma_s^{(+)} \rdot \non
&&
\times \ld \frac{1}{\big[\beta W(0)\big]^{1/2}} \frac{\tilde h}{\big(\tilde h^2+h_\text{c}^2\big)^{1/2}} \right], \non
&&
\frac{\partial}{\partial \beta\mu} \left[ e_0^{(+)} \tilde h \lp \tilde h^2 + h_\text{c}^2 \rp^{\frac{d-2}{2(d+2)}} - e_2^{(+)} \lp \tilde h^2 + h_\text{c}^2 \rp^{\frac{d}{d+2}}
\right] = \lp \tilde h^2 + h_\text{c}^2 \rp^{\frac{d-2}{2(d+2)}}  \non
&&
\times \left[ e_0^{(+)} \frac{1}{\big[\beta W(0)\big]^{1/2}} \lp 1+ \frac{d-2}{d+2} \frac{\tilde h^2}{\tilde h^2 + h_\text{c}^2} \rp - \frac{2d}{d+2} e_2^{(+)} \frac{1}{\big[\beta W(0)\big]^{1/2}} \frac{\tilde h}{\big(\tilde h^2 + h_\text{c}^2\big)^{1/2}}  \rdot\non
&&
- \ld \lp \tilde h^2 + h_\text{c}^2 \rp^{\frac{1}{2}} \frac{\partial e_2^{(+)}}{\partial \beta\mu}\right].
\eea
Note that during the calculation of the latter formula \eqref{3d6fa}, the derivatives of $\sigma_0$ with respect to $\beta\mu$ give the expression which coincide with the condition \eqref{2d28fa}. Therefore, the correspondent terms compensate each other. That is why, in the calculation scheme described above, the quantity $\sigma_0$ is considered not to be a function of chemical potential. The derivative ${\partial e_2^{(+)}}/{\partial\beta\mu}$ is as follows:
\be\label{3d7fa}
\frac{\partial e_2^{(+)}}{\partial\beta\mu} = \frac{1}{2} \sigma_0^2 s^{-3} \left( \frac{\partial r_{n_p+2}}{\partial\beta\mu} + \frac{1}{12} \sigma_0^2 \frac{\partial u_{n_p+2}}{\partial\beta\mu}\right),
\ee
where
\bea\label{3d8fa}
&&
\frac{\partial r_{n_p+2}}{\partial\beta\mu} = \frac{1}{\big[\beta W(0)\big]^{1/2}}  q E_2 \frac{\partial H_\text{c}}{\partial \tilde h}\,, \non
&&
\frac{\partial u_{n_p+2}}{\partial\beta\mu} = \frac{1}{\big[\beta W(0)\big]^{1/2}}  u^* \Phi_q E_2 \frac{\partial H_\text{c}}{\partial \tilde h}\,.
\eea
The expression for $H_\text{c}$ [see \eqref{2d10fa}] is applicable to find the derivative of $H_\text{c}$ with respect to $\tilde h$
\be\label{3d9fa}
\frac{\partial H_\text{c}}{\partial \tilde h} = - \frac{H_\text{c}}{p_0} \frac{\tilde h}{\tilde h^2 + h_\text{c}^2}\,.
\ee

\section{ Calculation of $\sigma_{00}^{(+)}$}
%\setcounter{equation}{0}
%\label{appB}

Let us provide the expressions of derivatives contained in the relation \eqref{3d13fa}. Taking into account the formulae \eqref{3d7fa}--\eqref{3d9fa}, one can derive the equality
\be\label{3d14fa}
\frac{\partial \text{e}^{(+)}_2}{\partial\beta\mu} = - \frac{1}{\big[\beta W(0)\big]^{1/2}} q_s \sigma_0^2 \left( 1 + \frac{q_l}{12} \sigma_0^2 \right) \big(\tilde h^2 + h_\text{c}^2\big)^{-1/2},
\ee
where the constants $q_s$ and $q_l$ are as follows:
\bea\label{3d15fa}
&&
q_s = \frac{E_2}{2p_0} q s^{-3} H_\text{c} \frac{\alpha}{\big(1+\alpha^2\big)^{1/2}}\,,\non
&&
q_l = \Phi_q u^* q^{-1},
\eea
 and $\alpha=\tilde h/h_\text{c}$ is the ratio of the renormalized chemical potential $\tilde h = M\big[\beta W(0)\big]^{1/2}$ $(M = \mu/W(0)-\tilde a_1)$ to the renormalized temperature $h_\text{c}=\tilde\tau^{(d+2)\nu/2}$ [$\tilde\tau = \tau(c_{11}/q)$]. The following is taken into account here:
\begin{equation}
\frac{\partial H_\text{c}}{\partial\beta\mu} = \frac{\partial H_\text{c}}{\partial\tilde h} \frac{\partial\tilde h}{\partial\beta\mu} = - \frac{1}{\big[\beta W(0)\big]^{1/2}} H_{\text{c}d} \lp \tilde h^2 + h_\text{c}^2 \rp^{-1/2}, \nonumber
\end{equation}
where
\begin{equation}
H_{\text{c}d} = \frac{H_\text{c}}{p_0} \frac{\alpha}{(1+\alpha^2)^{1/2}}\,. \nonumber
\end{equation}
The explicit expression of the derivative
\be\label{3d16fa}
\frac{\partial \gamma_s^{(+)}}{\partial\beta\mu} = \frac{\partial f_{n_p+1}}{\partial\beta\mu} - \frac{\partial \bar\gamma^{(+)}}{\partial\beta\mu} + s^{-3} \frac{\partial f_G}{\partial\beta\mu}
\ee
can be found by deriving the expressions of its each term.

Let us continue with the derivative of $f_{n_p+1}$ \eqref{2d18fa} with respect to the chemical potential. Note that
\bea\label{3d17fa}
&&
\frac{\partial y_{n_p+m}}{\partial\beta\mu} = y_{n_p+m} r_{p+m}\frac{\partial x_{n_p+m}}{\partial\beta\mu}, \quad r_{p+m}  = \frac{U'(x_{n_p+m})}{U(x_{n_p+m})} - \frac{1}{2} \frac{\varphi'(x_{n_p+m})}{\varphi(x_{n_p+m})}\,,\non
&&
U'(x_{n_p+m}) = \frac{1}{2} U^2(x_{n_p+m}) + x_{n_p+m} U(x_{n_p+m}) - 1\,, \non
&&
\varphi'(x_{n_p+m}) = 6 U'(x_{n_p+m}) U(x_{n_p+m}) + 2 U(x_{n_p+m}) + 2 x_{n_p+m} U'(x_{n_p+m})\,,
\eea
moreover, for $x_{n_p+m}$ \eqref{2d19fa}, we have
\be\label{3d18fa}
\frac{\partial x_{n_p+m}}{\partial\beta\mu} = \frac{1}{\big[\beta W(0)\big]^{1/2}} g_{p+m} \big( \tilde h^2 + h_\text{c}^2 \big)^{-1/2}.
\ee
Here,
\be\label{3d19fa}
g_{p+m} = - \bar x H_{\text{c}d} E_2^{m-1} \big( 1 + \Phi_q H_\text{c} E_2^{m-1} \big)^{-1/2} \left[ 1 - \frac{\Phi_q}{2} H_\text{c} E_2^{m-1} \big( 1 + \Phi_q H_\text{c} E_2^{m-1}\big)^{-1} \right].
\ee
Taking into account the equalities represented above,
\be\label{3d20fa}
\frac{\partial f_{n_p+1}}{\partial\beta\mu} = \frac{1}{\big[\beta W(0)\big]^{1/2}} f_p \big( \tilde h^2 + h_\text{c}^2 \big)^{-1/2},
\ee
where
\be\label{3d21fa}
f_p = \frac{1}{2} r_p g_p \big( 1 - 9 / y^2_{n_p}\big) - \frac{1}{2} g_{p+1} U(x_{n_p+1}).
\ee
The derivative of $\bar\gamma^{(+)}$ [see \eqref{2d7fa}] with respect to the chemical potential is of the form
\be\label{3d22fa}
\frac{\partial \bar\gamma^{(+)}}{\partial\beta\mu} = - \frac{1}{\big[\beta W(0)\big]^{1/2}} \gamma_p \big( \tilde h^2 + h_\text{c}^2 \big)^{-1/2}.
\ee
Here,
\be\label{3d23fa}
\gamma_p = H_{\text{c}d} \lp \bar\gamma_2 + 2 \bar\gamma_3 H_\text{c} \rp.
\ee
Calculating the derivative of $f_G$ \eqref{2d41fa} with respect to the chemical potential, one should take into account that the shift $\sigma_0$ is a function of chemical potential. To derive $\partial \sigma_0/\partial\beta\mu$, let us use the equality \eqref{2d28fa}, which makes possible to obtain
\be\label{3d24fa}
\frac{\partial \sigma_0}{\partial\beta\mu} = \frac{1}{\big[\beta W(0)\big]^{1/2}} g_\sigma \big( \tilde h^2 + h_\text{c}^2 \big)^{-1/2},
\ee
where
\be\label{3d25fa}
g_\sigma = \frac{s^{5/2}}{r_R} \frac{1}{1+\alpha^2} + \frac{\sigma_0}{r_R} H_{\text{c}d} q E_2 \lp 1 + \frac{q_l}{6} \sigma_0^2 \rp.
\ee
The derivative of $r_R$ is
\be\label{3d26fa}
\frac{\partial r_R}{\partial\beta\mu} = \frac{1}{\big[\beta W(0)\big]^{1/2}} g_R \big( \tilde h^2 + h_\text{c}^2 \big)^{-1/2},
\ee
where
\be\label{3d27fa}
g_R = -q E_2 H_{\text{c}d} \lp 1 + \frac{1}{2} q_l \sigma_0^2 \rp + u_{n_p+2} g_\sigma \sigma_0\,,
\ee
and the derivative of $a$ from \eqref{2d37fa} is
\be\label{3d28fa}
\frac{\partial a}{\partial\beta\mu} = \frac{1}{\big[\beta W(0)\big]^{1/2}} g_a \big( \tilde h^2 + h_\text{c}^2 \big)^{-1/2}, \quad g_a = - \frac{a g_R}{2r_R}.
\ee
The derivative of the quantity $f''_G$ \eqref{2d43fa} contained in $f_G$ [see \eqref{2d41fa}] is as follows:
\be\label{3d29fa}
\frac{\partial f''_G}{\partial\beta\mu} = \frac{1}{\big[\beta W(0)\big]^{1/2}} g_a a_g \big( \tilde h^2 + h_\text{c}^2 \big)^{-1/2}.
\ee
Here,
\be\label{3d30fa}
a_g = \frac{2a}{1+a^2} - \frac{4}{a^3} + \frac{6}{a^4} \arctan a - \frac{2}{a^3} \frac{1}{1+a^2}.
\ee
Taking into account the expressions represented above, we come to the following expression
\be\label{3d31fa}
\frac{\partial f_G}{\partial\beta\mu} = \frac{1}{\big[\beta W(0)\big]^{1/2}} f_{gv} \big( \tilde h^2 + h_\text{c}^2 \big)^{-1/2},
\ee
where
\be\label{3d32fa}
f_{gv} = - \frac{1}{4} \frac{u^* \Phi_q}{u_{n_p+1}} H_{\text{c}d} - \frac{1}{2} \bigg( \frac{g_R}{r_R} + g_a a_g \bigg) + g_{p+1} \bigg[ \frac{3}{4} \frac{r_{p+1}}{y_{n_p+1}^2} - \frac{1}{2} \frac{U'(x_{n_p+1})}{U(x_{n_p+1})} \bigg].
\ee

Summing up the terms \eqref{3d20fa}, \eqref{3d22fa} and \eqref{3d31fa}, we obtain the derivative of $\gamma_s^{(+)}$ with respect to the chemical potential based on \eqref{3d16fa}
\be\label{3d33fa}
\frac{\partial \gamma_s^{(+)}}{\partial\beta\mu} = \frac{1}{\big[\beta W(0)\big]^{1/2}} f_{\gamma_1} \big( \tilde h^2 + h_\text{c}^2 \big)^{-1/2}.
\ee
Here,
\be\label{3d34fa}
f_{\gamma_1} = f_p + \gamma_p + f_{gv} / s^3.
\ee
The quantities $f_p$, $\gamma_p$ and $f_{gv}$ are functions of $\alpha$ only.

The obtained expressions \eqref{3d12fa}, \eqref{3d13fa}, \eqref{3d14fa} and \eqref{3d33fa} are helpful in calculating the coefficient $\sigma_{00}^{(+)}$ in the equality \eqref{3d10fa}. Therefore,
\be\label{3d35fa}
\sigma_{00}^{(+)} = e_0^{(+)} \frac{1}{\big[\beta W(0)\big]^{1/2}} \lp 1 + \frac{d-2}{d+2} \frac{\alpha^2}{1+\alpha^2} \rp + e_{00}^{(+)}
\frac{\alpha}{\big(1+\alpha^2\big)^{1/2}} + e_{02}^{(+)}\,,
\ee
where
\be\label{3d36fa}
e_{02}^{(+)} = e_{01}^{(+)} \big( \tilde h^2 + h_\text{c}^2 \big)^{1/2} = \frac{1}{\big[\beta W(0)\big]^{1/2}} \left[ f_{\gamma_1} + q_s\sigma_0^2 \big( 1 + q_l\sigma_0^2/12 \big) \right].
\vspace{-1mm}
\ee
Obviously, $\sigma_{00}^{(+)}$ is not a function of either temperature or of chemical potential. It is a function of their ratio
\[
\alpha = \frac{\tilde h}{h_\text{c}} = \big[\beta W(0)\big]^{1/2} \lp \frac{q}{c_{11}}\rp^{p_0} \alpha_0.
\]
The multiplier
\[
\alpha_0 = \frac{\mu / W(0) - \tilde a_1}{\tau^{p_0}}
\]
is defined by the ratio of initial $\mu$ and $\tau$.

%
%% If you have problems with typesetting in ukrainian uncomment lines below.
%
%  \lastpage
%  \end{document}
\newpage
\ukrainianpart

\title{Рівняння стану коміркової моделі плину в надкритичній області}
\author{М.П. Козловський, І.В. Пилюк, О.А. Добуш}
\address{Інститут фізики конденсованих систем НАН України, вул. Свєнціцького, 1, 79011 Львів, Україна}

\makeukrtitle

\begin{abstract}
\tolerance=3000%
Використовуючи множину колективних змінних та перетворення ренормалізаційної групи, розвинено аналітичний метод розрахунку рівняння стану коміркової моделі плину в області вище критичної температури ($T \geqslant T_\text{c}$).
Математичний опис з урахуванням негаусових флуктуацій параметра порядку виконано в околі критичної точки на основі
моделі $\rho^4$.
Запропонований метод розрахунку великої статистичної суми дозволяє отримати, крім універсальних величин, зокрема, критичних показників кореляційної довжини, рівняння для критичної температури моделі плину.
Побудовано криві ізотермічної стисливості як функції густини.
Також зображено лінію екстремумів стисливості у надкритичній області.

\keywords коміркова модель плину, критичні показники, рівняння стану, надкритична область

\end{abstract}

\end{document}